\documentclass[11pt]{article}
\pdfoutput=1

\usepackage{jheppub}
\usepackage[utf8]{inputenc}
\usepackage{booktabs}
\usepackage{graphicx}
\usepackage{bbold}
\usepackage{subcaption}
\usepackage{mathtools}
\usepackage{amsmath}
\usepackage{kantlipsum}
\allowdisplaybreaks
\usepackage{autobreak}

\numberwithin{equation}{section}

\newcommand{\nn}{\nonumber}

\newcommand\beq{\begin{equation}}
\newcommand\eeq{\end{equation}}
\newcommand\beal{\begin{aligned}}
\newcommand\eeal{\end{aligned}}
\newcommand\bea{\begin{eqnarray}}
\newcommand\eea{\end{eqnarray}}

\newcommand\dd{{\mathrm d}}
\usepackage{xcolor}

\newcommand{\bk}{{\boldsymbol k}}

\newcommand{\bp}{{\boldsymbol p}}

\newcommand{\bL}{{\boldsymbol L}}

\newcommand{\br}{{\boldsymbol r}}

\newcommand\cE{\mathcal{E}}

\newcommand\cS{\mathcal{S}}

\makeatletter
\newcommand{\Biggg}{\bBigg@{3.5}}
\makeatother

\definecolor{1PN}{rgb}{0.00,0.28,0.73}
\definecolor{2PN}{rgb}{0.40,0.00,0.80}
\definecolor{3PN}{rgb}{0.70,0.00,0.23}
\definecolor{4PN}{rgb}{0.70,0.35,0.00}
\definecolor{5PN}{rgb}{0.33,0.50,0.00}
\definecolor{6PN}{rgb}{0.00,0.40,0.33}
\definecolor{7PN}{rgb}{0.80,0.2,0.00}

\begin{document}
\preprint{DESY\, 21-212\\\phantom{~}}
\title{\center From Boundary Data to Bound States III:\\ [0.2cm] Radiative Effects}
\author{\large \,\,\,\,Gihyuk Cho,\,}
\author{\large Gregor K\"alin}
\author{\large and Rafael A. Porto}
\affiliation{Deutsches Elektronen-Synchrotron DESY, Notkestr. 85, 22607 Hamburg, Germany}

\emailAdd{gihyuk.cho@desy.de,\,gregor.kaelin@desy.de,\,rafael.porto@desy.de}
\abstract{We extend the {\it boundary-to-bound} (B2B) correspondence to incorporate radiative as well as conservative radiation-reaction effects. We start by deriving a map between the total change in observables due to gravitational wave emission during hyperbolic-like motion and in one period of an elliptic-like orbit, which is valid in the adiabatic expansion for non-spinning as well as aligned-spin configurations. We also discuss the inverse problem of extracting the associated fluxes from scattering data. Afterwards we demonstrate, to all orders in the Post-Minkowskian expansion, the link between the radiated energy and the ultraviolet pole in the radial action in dimensional regularization due to tail effects. This implies, as expected, that the B2B correspondence for the conservative sector remains unchanged for local-in-time radiation-reaction tail effects with generic orbits. As~a~side product, this allows us to read off the energy flux from the associated pole in the tail Hamiltonian. We~show that the B2B map also holds for non-local-in-time terms, but only in the {\it large-eccentricity} limit.  Remarkably, we find that all of the trademark logarithmic contributions to the radial action map unscathed between generic unbound and bound motion. However, unlike logarithms, other terms due to non-local effects do not transition smoothly to {\it quasi-circular}~orbits.  We~conclude with a discussion on these non-local pieces. 
Several checks of the  B2B dictionary are displayed using state-of-the-art knowledge in Post-Newtonian/Minkowskian theory.} 

\maketitle
\newpage

\section{Introduction} \label{sec:sum}
Motivated by the `on-shell' nature of scattering processes, a  {\it boundary-to-bound} (B2B) correspondence between observables, such as the deflection angle and periastron advance, was introduced in \cite{paper1,paper2} for the (strictly) conservative sector. The B2B map connects hyperbolic- to elliptic-like orbits through analytic continuation of the radial action without resorting to gauge-dependent objects like the Hamiltonian. With the goal of using Post-Minkowskian (PM) scattering data --- obtained through (quantum) amplitude-based, e.g.~\cite{cheung,Guevara:2018wpp,donal,donalvines,bohr,zvi1,zvi2,zvispin,soloncheung,Chung:2020rrz,andres2,4pmzvi,DiVecchia:2021bdo,Cristofoli:2021vyo,parra2,Bautista:2021inx}, and (classical) EFT-based, e.g.~\cite{Damour:2016gwp,Goldberger:2016iau,Damour:2019lcq,pmeft,3pmeft,tidaleft,pmefts,4pmeft,4pmtail,janmogul,janmogul2,max1,Jakobsen:2021lvp,max2,Bini:2021gat}, methodologies ---  to construct high-precision Post-Newtonian (PN) waveform models, e.g. \cite{Antonelli:2019ytb}, the purpose of this paper is to extend the B2B dictionary to the radiation sector.  Radiative effects in the two-body dynamics come in two flavors. Firstly, there is the obvious loss of energy and angular momentum due to gravitational wave (GW) emission responsible for the celebrated change in the orbital parameters. There are, however, also `conservative' radiation-reaction contributions to, for instance, the binding energy. These arise due to non-linear gravitational effects, such as the GWs getting trapped in the near zone through scattering off of the background  geometry, so-called {\it tail} effects, see e.g. \cite{Blanchet:1987wq,tail2}, or off of the waves emitted at an earlier time, so-called the (non-linear) {\it memory}.\footnote{We will concentrate mostly on tail terms in this paper. We briefly discuss memory corrections in \S\ref{disc}.} We~study~here the B2B map for both dissipative and conservative radiation-reaction effects.\vskip 4pt

In principle, the instantaneous dynamics due to dissipation produces various transient phenomena associated with radiation-reaction forces which cannot be captured by the conservative equations of motion, e.g. \cite{chadira,natalia1,natalia2}. However, provided the adiabatic expansion holds, we can evaluate the total loss of energy and angular momentum by integrating the (averaged) instantaneous flux using the unperturbed (conservative) solution. Here is where the original B2B dictionary \cite{paper1,paper2} --- linking the point of closest approach in a scattering process to the endpoints of elliptic-like orbits via analytic continuation in the angular momentum and binding energy --- becomes extremely powerful.  After taking into account the parity properties under $J \to -J$ of the associated fluxes, we show the following relationships,
\beq
\Delta E_{\rm ell}(J) = \Delta E_{\rm hyp}(J) -\Delta E_{\rm hyp}(-J)\,, \quad\quad 
\Delta J_{\rm ell}(J) = \Delta J_{\rm hyp}(J) + \Delta J_{\rm hyp}(-J) \,,\nn
\eeq
hold between the total radiated (source) energy and angular momentum in a scattering process and one orbit of elliptic-like motion, valid for non-spinning or aligned-spin configurations.  As~we shall see, the analytic continuation also holds in terms of the eccentricity parameter, which will be useful when performing the B2B map between radiative observables. \vskip 4pt 

Although the above relations already uncover a non-trivial link between radiation effects for unbound and bound states, to compute the GW phase for generic orbits requires the (averaged) instantaneous flux. From the point of view of scattering, the inverse problem then turns into obtaining the flux from the knowledge of the total radiated energy. Following similar steps as for the Firsov representation discussed in \cite{paper1,paper2}, we show here how to reconstruct the GW flux in an isotropic gauge from the PM expansion of the radiated~energy.\vskip 4pt 

In order to complete all the necessary ingredients for accurate waveform modeling we must also incorporate conservative corrections due to hereditary effects. As it turns out, these come in different forms since not only they introduce the standard local-in-time interactions, but are also responsible for non-local-in-time effects \cite{Blanchet:1987wq}. The latter are a trademark of tail-type contributions and do not appear with memory terms. In combination with the existence of spurious infrared/ultraviolet divergences in intermediate computations from the near (potential) and far (radiation) zones, the time non-locality induces a breakdown of the near/far zone expansion that caused the introduction of several (apparent) {\it ambiguities} in  `traditional' PN computations \cite{blanchet,Schafer:2018kuf,Damour:2014jta,Jaranowski:2015lha,Bernard:2017bvn,Marchand:2017pir}.  These issues, however, are naturally handled within the EFT approach \cite{nrgr,walterLH,foffa,iragrg,grg13,review}, resulting in completely unambiguous answers \cite{tail,apparent,nrgr4pn1,nrgr4pn2}. Yet, because of a variety of contributions and mixing between zones, the EFT split into potential and radiation modes can be further decomposed into (generalized) local- and non-local-in-time effects. The latter separation, which we adopt in this paper, is advantageous when it comes to applying the B2B dictionary as well as incorporating the true logarithmic corrections to the binary dynamics.\vskip 4pt

In order to establish the extension of the B2B correspondence, we start by demonstrating the universal connection between the energy flux, ${\cal F}_E$, and the ultraviolet-divergent part of the tail contribution to the Hamiltonian, $H_{\rm tail}$, due to radiation modes \cite{4pmeft,4pmtail},
\beq
\frac{H_{\rm tail}(r,\bp^2)}{GE} = \frac{{\cal F}_E(r,\bp^2)}{(d-4)_{\rm UV}} +\cdots \,, \nn
\eeq
which cancels out against an infrared pole arising from the potential region \cite{apparent}. From~here, and the connection between the effective and radial actions, we conclude that the B2B dictionary remains valid for local-in-time effects. The fact that the latter can be mapped from unbound to bound motion through the original B2B dictionary is not surprising. After all, if we shuffle all of the logarithms into the non-local part the remaining pieces can be parameterized using the standard PM expansion for the center-of-mass momentum which, using the manipulations in \cite{paper1,paper2}, yield the original B2B relations. What is instructive, however, is to notice that, due to the universal connection between the flux and the tail Hamiltonian, or likewise the radiated energy and radial action, the B2B correspondence in the dissipative sector is intimately linked to the B2B map for conservative local-in-time~effects.\vskip 4pt

The situation changes dramatically for non-local-in time terms. While we can, once again, resort to an adiabatic expansion in which the hereditary integral is computed over the conservative motion, the resulting dynamical equations vary significantly between hyperbolic-like motion and generic bound orbits. Nonetheless, we find that the original B2B correspondence applies \cite{paper1,paper2}
\beq
\cS_{r, \rm ell}(J) = \cS_{r, \rm hyp}(J) - \cS_{r, \rm hyp}(-J)\nn\,,
\eeq
for the full radial action, but only in the limit of large angular momentum, also known as the {\it large-eccentricity} limit. \\

Unfortunately, the resulting coefficients in the $1/J$ expansion do not capture generic terms associated with non-local-in-time effects for e.g. quasi-circular motion. Nevertheless, not only can we map all of the local-in-time contributions in this fashion, remarkably the large-eccentricity limit readily accounts for all the leading logarithmic corrections in $\cE$, the (reduced) binding energy, which take on the universal form
\beq
\cS_{r, \rm hyp/ell}^{\log} = -\frac{GE}{2\pi} \Delta E_{\rm hyp/ell} \log |\cE|\,,\nn
\eeq
and may be obtained through the original B2B (bound) radial action via the scattering angle~\cite{paper1,paper2}.
 Moreover, we also find the center-of-mass momentum can be reconstructed in terms of an (effectively local) PM expansion in $G/r$, which correctly incorporates both local-in-time as well as non-local-in-time effects proportional to logarithms of the binding energy. In~combination with Firsov's representation, this allows us to read off all of the associated coefficients in the $1/J$ expansion of the radial action to a given PN order~\cite{paper1,paper2}. At the same time, it illustrates how --- unlike the intermediate potential-only results --- the spurious factors of $\log r$ in the center-of-mass momentum \cite{4pmeft} or Hamiltonian \cite{4pmzvi} do not appear in the complete (ambiguity-free) dynamics \cite{4pmtail}.\vskip 4pt  
 
 After demonstrating the existence of an extended B2B correspondence in the radiative sector,\footnote{An intriguing  connection resembling the B2B map was recently discussed in \cite{bianchi}. It would be interesting to explore whether it also extends to the radiative sector.} we dedicate the penultimate section to various explicit checks of the B2B map using the state-of-the-art in the PN/PM expansion both for non-spinning and aligned-spin configurations. To the extent of our knowledge, the computations of the total radiated energy and angular momentum including spin effects using the fluxes obtained in \cite{Cho:2021mqw} --- which we display here in their full glory --- are presented for the first time. While our results for the conservative sector are completely generic, we restrict ourselves to the well understood local- and non-local-in-time contributions at 4PN order derived through different methodologies in \cite{tail,apparent,Damour:2014jta,Jaranowski:2015lha,Bernard:2017bvn,Marchand:2017pir,nrgr4pn1,nrgr4pn2}, except for logarithmic terms which we incorporate to 6PN order using the results in \cite{binidam2}. In addition, we have explicitly checked the B2B map applies to the contributions at 5PN reported in \cite{5pnfin2} as well as  the intermediate results in \cite{binidam1}. We~will make a few comments at the end regarding conservative (local-in-time) memory terms.\vskip 4pt The final section of this paper is devoted to a discussion on plausible ways to map generic non-local-in-time effects from scattering processes to generic bound states.  Throughout this paper we follow the conventions and notation of \cite{paper1,paper2}, to which we refer to reader for further details. We also provide an ancillary file collecting the main expressions presented here.

\newpage
\section{Orbital elements}
\begin{figure}[t!] 
\begin{center}
\includegraphics[width=0.65\textwidth]{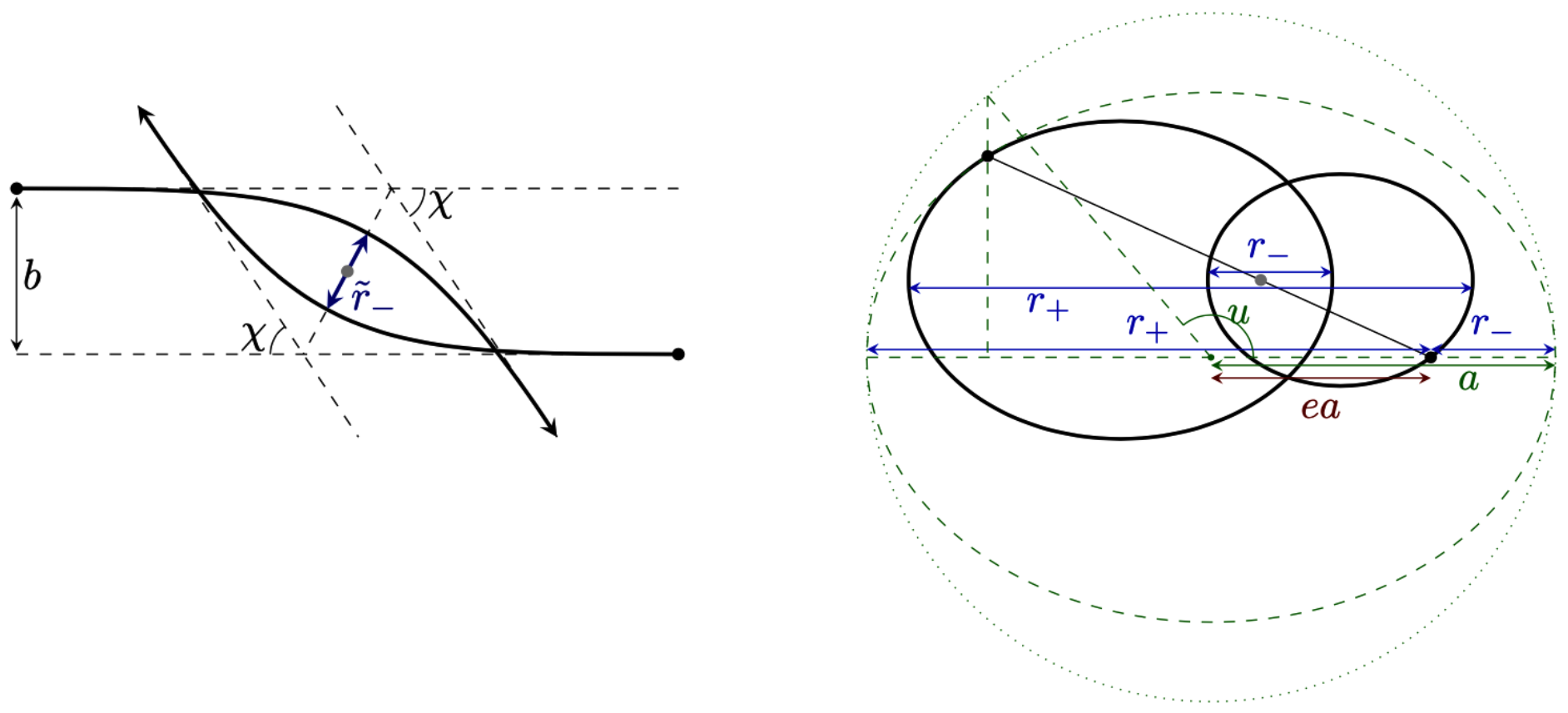} 
      \caption{The geometry for unbound and bound motion. See the text for definition of the various variables.} 
      \label{fig1}
            \vspace{-0.4cm}
            \end{center}
\end{figure}

One of the key ideas behind the B2B dictionary is the link between the endpoints of elliptic-like motion and closest approach in hyperbolic encounters, see Fig.~\ref{fig1}. Here we summarize the correspondence and add also a few additional elements which play an important role later on when performing the analytic continuation between unbound and bound motion. 

\subsection{$J \to -J$}
In the original B2B map in \cite{paper1,paper2} for observables in the conservative sector, the endpoints of a bound orbit, $0 < r_- < r_+$, were shown to be related to the distance of closest approach in a scattering event, $\tilde r_-$, via analytic continuation,  
 \bea
 r_-(J,\cE) &=&  \tilde r_- (J ,\cE)\,,\label{rpm}\\
 r_+ (J,\cE) &=&  \tilde r_- (-J,\cE)\,, \nn
 \eea
with positive (total) angular momentum, $J>0$, and negative binding energy, $\cE \equiv \tfrac{E-M}{M\nu} <0$, respectively, where $M=m_1+m_2$ is the total mass, $E$ the total energy, and $\nu$ the symmetric-mass-ratio. The above relationships are a direct consequence of Firsov's representation for the momentum of the incoming particles in the center-of-mass frame, \cite{firsov}  
 \beq
 \label{firsov1}
\bar\bp^2 = \exp\left[ \frac{2}{\pi} \int_{r|\bar\bp|}^\infty \frac{ \chi (\tilde b_c)\dd\tilde b_c}{\sqrt{\tilde b_c^2-r^2\bar\bp^2}}\right] =1+\sum_{i=1}^\infty f_i (\cE,L,a_\pm)\left(\frac{GM}{r}\right)^i \,, 
\eeq 
where $\bar\bp^2 \equiv \bp^2/p^2_\infty$ with $p_\infty$ the incoming momentum at infinity, $\chi(\tilde b_c)$ is the scattering angle, $a_\pm = a_1 \pm a_2$ with $a_i$ the (mass-rescaled) spin parameters, and $b_c \equiv L/p_\infty$. The (canonical) orbital angular momentum, $L$, obeys the relation  \cite{justin1}
\beq
L = p_\infty b + M \frac{\Gamma-1}{2}\left(a_+ - \frac{\Delta }{\Gamma} a_-\right)\,,
\eeq
with $b$ the (covariant) impact parameter. We have also introduced the notation $\Gamma\equiv E/M$, $\Delta \equiv \sqrt{1-4\nu}$. From here we obtain
\beq
\label{firsov2}
\tilde r_- =  b_c \,\exp\left[ -\frac{1}{\pi} \int_{b_c}^\infty \frac{\chi(\tilde b_c)\dd\tilde b_c}{\sqrt{\tilde b_c^2-b_c^2}}\right]\,,
\eeq
for the (real and positive) root of the vanishing radial momentum at the point of closest proximity, 
\beq 
\bar p_r^2(\tilde r_-) =  \bar\bp^2(\tilde r_-) - b_c^2/{\tilde r_-}^2 = 0\,.
\eeq
\vskip 4pt  It is straightforward to show a relationship similar to the one implied by \eqref{rpm} also applies for the other (unphysical) root, $\tilde r_+ < 0$, of the unbound motion; namely
\beq
\tilde r_+(J,\cE) =  \tilde r_- (-J,\cE)\,, \label{rpm2}
\eeq  
from positive (total) angular momentum $J>0$, but this time having both positive binding energy, $\cE>0$.
\subsection{$e \to -e$} 
From the roots of the radial momentum we can construct the parameters describing the eccentricity of the orbit,
\beq
\begin{aligned}
 e&= \frac{r_+ - r_-}{r_++r_-} \quad \quad {\rm (ellipse)}\,,\\
\tilde e &= \frac{\tilde r_+ - \tilde r_-}{\tilde r_++\tilde r_-} \quad \quad {\rm (hyperbola)}\,.
\end{aligned}
\eeq
From the conditions in \eqref{rpm} and \eqref{rpm2} we notice that $J \to -J$ implies  $e \to -  e$, and similarly $\tilde e \to - \tilde e$. This means, for instance, the B2B relationship between scattering angle and periastron advance \cite{paper2} 
\beq
\Delta \Phi ( J,\cE) =  \chi( J,\cE) + \chi(- J,\cE)\,, \label{Phij}
\eeq
may also be written as
\beq
\Delta \Phi ( e,\cE) =  \chi( e,\cE) + \chi(- e,\cE)\,, \label{Phie}
\eeq
analytically continued to $e<1$ and negative binding energies.\footnote{We thank M. van de Meent for demonstrating the  exact validity of \eqref{Phij}-\eqref{Phie} for the case of a test body in a Schwarzschild background (unpublished).}
The map involving eccentricities will prove to be useful when studying radiative observables next. 

\section{Radiative observables}\label{sec:rad}
In this section we focus on the total change induced on observables due to GW radiation in a scattering process and one period of elliptic-like motion. In all cases we will assume the validity of the adiabatic expansion; namely, that we can evaluate the total change by integrating the (averaged) fluxes over the solution of the conservative  dynamical equations. This will allow us to use the connection between the orbital elements described above. Moreover, we will concentrate here either on non-spinning or aligned-spin configurations, such that motion remains in a plane orthogonal to the total angular momentum. 

\subsection{$J$-parity} \label{sec:rad1}
Let us consider the (averaged) flux, ${\cal F}_{\cal O}(r,\cE, L,a_\pm)$, associated with an observable quantity. The total change may be then written as follows
\beq
\begin{aligned}
\Delta {\cal O}_{\rm ell} = \int_0^{T_{\rm orb}} {\cal F}_{\cal O}(r,\cE,L,a_\pm) dt &= 2\int_{r_-}^{r_+}{\cal F}_{\cal O}(r,\cE,L,a_\pm)\frac{dr}{\dot r}\,, \\
\Delta {\cal O}_{\rm hyp}=\int_{-\infty}^{+\infty} {\cal F}_{\cal O}(r,\cE,L,a_\pm) dt &= 2\int_{\tilde r_-}^{\infty}{\cal F}_{\cal O}(r,\cE,L,a_\pm)\frac{dr}{\dot r}\,,
\end{aligned}
\eeq
 for elliptic- and hyperbolic-like motion, respectively. Assuming the validity of an adiabatic approximation, and using the conservative equations of motion in a quasi-isotropic gauge\footnote{In principle there is also a correction to $\dot\br$ proportional to $\br \times \tfrac{\partial H}{\partial \bL}$. However, it does not contribute to~$\dot r$.} 
\beq 
\dot r = 2 {\partial H \over \partial \bp^2} \, p_r \,, \label{dotr}
\eeq
with $H$ the Hamiltonian, as well as (see \eqref{firsov1})
\beq
p_r^2 = \bp^2-J^2/r^2\,,
\eeq
we arrive at
\beq
\begin{aligned}
\Delta{\cal O}_{\rm ell}  &= \int_{r_-}^{r_+}\left(\partial H \over \partial \bp^2\right)^{-1}\frac{{\cal F}_{\cal O}(r,\cE,L,a_\pm)}{\sqrt{\bp^2(r,\cE,L,a_\pm)-J^2/r^2}}dr \,, \\
\Delta {\cal O}_{\rm hyp} &= \int_{\tilde r_-}^{\infty}\left(\partial H \over \partial \bp^2\right)^{-1}\frac{{\cal F}_{\cal O}(r,\cE,L,a_\pm)}{\sqrt{\bp^2(r,\cE,L,a_\pm)-J^2/r^2}}dr\,.
\end{aligned}
\eeq
Since the Hamiltonian is a function of $r$, $\bp^2$, $a_\pm^2$ , and $L a_\pm$, the extension of the B2B dictionary hinges upon the properties under a ``$J$-parity"  transformation: $L \to -L$ and $a_\pm \to -a_\pm$, of the given fluxes
\beq
{\cal F}_{\cal O}(r,\cE,-J) = \sigma_{\cal O}\, {\cal F}_{\cal O}(r,\cE,J)\,,
\eeq
with $\sigma_{\cal O} = \pm 1$. By performing the same {\it loop around infinity} as in \eqref{Phij}, we find
\beq
\Delta{\cal O}_{\rm ell} (J,\cE) = \Delta {\cal O}_{\rm hyp}(J,\cE) -\sigma_{\cal O} \Delta {\cal O}_{\rm hyp}(-J,\cE)\,,
\eeq  
which, as we discussed earlier, can also be written as
\beq
\Delta {\cal O}_{\rm ell}(e,\cE) = \Delta {\cal O}_{\rm hyp}(e,\cE) -\sigma_{\cal O} \Delta {\cal O}_{\rm hyp}(-e,\cE)\,.
\eeq  
These manipulations allow us to relate the total change for several observable quantities computed in unbound and bound orbits.
\subsection{Energy}\label{sec:rad2}
The case of the total radiated energy is the simplest. Firstly, we notice that in an quasi-isotropic gauge we have 
\beq
{\cal F}_{E}(r,\cE,-J) = +{\cal F}_{E}(r,\cE,J)\,,
\eeq
for the energy flux, almost trivially since it can be written as a function of $r$, $\bp^2$, $a_\pm^2$ and $La_\pm$.\footnote{The existence of this gauge is also connected to the link between the tail Hamiltonian and energy flux.} This is not surprising since we expect the flux to be a scalar under $J$-parity. Following our previous reasoning we then arrive at \cite{4pmeft}
\beq
\begin{aligned}
\Delta E_{\rm ell}(J,\cE) &= \Delta E_{\rm hyp} (J,\cE) - \Delta E_{\rm hyp}(-J,\cE)\,,\label{b2ben}\\
\Delta E_{\rm ell}(e,\cE) &= \Delta E_{\rm hyp} (e,\cE) - \Delta E_{\rm hyp}(-e,\cE)\,,
\end{aligned} 
\eeq
which was previously observed to hold in \cite{binidam2} (for the $J \to -J$ map) with the knowledge of the PN expansion at the time. We show here that, as expected, it continues to hold for the recently derived results at 3PN order \cite{cho3pn}.

\subsection{Angular momentum}\label{sec:rad3}

We can also apply the map to the angular momentum. Being a pseudo-vector, its flux must transform under $J$-parity in the same fashion as the angular momentum itself,
\beq
{\cal F}_J(r,\cE,-J) = -{\cal F}_J(r,\cE,J)\,.
\eeq
Therefore, the correspondence between elliptic- and hyperbolic-like motion becomes
\beq
\begin{aligned}
\Delta J_{\rm ell}(J,\cE) &= \Delta J_{\rm hyp} (J,\cE) + \Delta J_{\rm hyp}(-J,\cE)\,,\label{b2bang}\\
\Delta J_{\rm ell}(e,\cE) &= \Delta J_{\rm hyp} (e,\cE) + \Delta J_{\rm hyp}(-e,\cE)\,,
\end{aligned} 
\eeq
similarly to that of the periastron advance and scattering angle. 

\subsection{Inverse problem}\label{inverse}

For the conservative sector, the Firsov representation in \eqref{firsov1} allows us to construct a (gauge-dependent) object to describe the dynamics from the knowledge of the scattering angle. The main idea is to invert the integral which defines the latter in terms of the radial momentum. A~similar representation can then be shown to exist for the GW flux in terms of the total radiated energy and angular momentum. For the sake of simplicity we consider the non-spinning case. Let us stare  at the expressions for the total radiated energy,
\beq
\begin{aligned}
\Delta { E}_{\rm hyp} &= \int_{\tilde r_-}^{\infty}\left(\partial H(r,\bp^2) \over \partial \bp^2\right)^{-1}\frac{{\cal F}_{ E}(r,\cE)}{\sqrt{\bp^2(r,\cE)-J^2/r^2}}dr  \label{firsovE0}
\\
\Delta { E}_{\rm ell} &= \int_{ r_-}^{r^+}\left(\partial H(r,\bp^2) \over \partial \bp^2\right)^{-1}\frac{{\cal F}_{ E}(r,\cE)}{\sqrt{\bp^2(r,\cE)-J^2/r^2}}dr
\end{aligned}
\eeq
 Since we are only concerned about the source energy, the associated flux may be expanded in powers of $GM/r$,
\beq
{\cal F}_{ E}(r,\cE) =\frac{M}{r} \sum_{n=0}^{\infty} {\cal F}^{(n)}_{E}(\cE) \left(\frac{G M}{r}\right)^{n+3}\,,\label{firsovE}
\eeq
with ${\cal F}^{(n)}_{\cal E}(\cE)$ (dimensionless) functions of the binding energy, and we chose the starting point of the sum in hindsight of the leading PN effects. From here, and the fact that the PM expansion of the Hamiltonian (and its derivatives) can be obtained via scattering data \cite{cheung,paper1,paper2}, we arrive at an integral involving only the radial motion and some functions of the binding energy. By performing the integration and expanding in $G/J$ we can then read off the ${\cal F}^{(n)}_{\cal E}(\cE)$ coefficients directly from the PM expansion of the total radiated energy
\beq
\begin{aligned}
\Delta { E}_{\rm hyp}(j,\cE) &=  \sum_{n=0}^{\infty} \Delta E_{\rm hyp}^{(n)}(\cE) \frac{1}{j^{n+3}}\,,\\
\Delta { E}_{\rm ell}(j,\cE) &=  \Delta { E}_{\rm hyp}(j,\cE) - \Delta { E}_{\rm hyp}(-j,\cE)\,,
\end{aligned}
\eeq
with $j \equiv J/(GM^2\nu)$. The final expressions, however, are rather cumbersome and not particularly illuminating (see ancillary file). Here we quote only the first two orders, obtained using the known PM results for the conservative sector~\cite{pmeft}, 
\beq
\begin{aligned}
\label{f12}
    M\pi\xi \,{\cal F}_E^{(0)} &= \frac{2 \Gamma  \nu \Delta E_\textrm{hyp}^{(0)}}{\left(\gamma ^2-1\right) }\,,\\
    M\pi\xi \, {\cal F}_E^{(1)} &= \frac{3 \pi  \Gamma ^2 \nu \Delta E_\textrm{hyp}^{(1)} }{4 \left(\gamma ^2-1\right)^{3/2}}-\frac{2 \Delta E_\textrm{hyp}^{(0)} \nu ^3}{\left(\gamma ^2-1\right)^2 \Gamma ^6 \xi ^2} \bigg[(\gamma -1)^3 \left(10 \gamma ^3-10 \gamma ^2-9 \gamma +5\right) \nu ^2\\
      &\quad\quad+4 \left(5 \gamma ^5-8 \gamma ^4+\gamma ^3+4 \gamma ^2-3 \gamma +1\right) \nu +8 \gamma ^4-4 \gamma ^2-1\bigg]\,,
  \end{aligned}
\eeq
with \beq \xi \equiv E_1E_2/E^2\,, \quad E_a \equiv \sqrt{p_\infty^2+m_a^2}\,,\quad p_\infty^2 = \frac{M \nu}{\Gamma} (\gamma^2-1)\,,\quad \gamma \equiv 1+\cE + \frac{\nu}{2} \cE^2 \,,\eeq
and provide an explicit derivation at 3PM in~\S\ref{secpnm}. A~similar reasoning applies to the angular momentum. As we shall see shortly, the coefficients in the PM expansion in \eqref{firsovE} can be also read off from the imprint of radiation-reaction effects in the conservative dynamics. 
 
 \section{Conservative radiation-reaction}
As it is well known hereditary back-reaction effects contribute to the conservative regime of the dynamics, e.g. \cite{tail,tail2}.  In this section we discuss the extension of the B2B dictionary to conservative radiation-reaction effects. 
  
\subsection{Universality}

\begin{figure}[t!] 
\begin{center}
\includegraphics[width=0.8\textwidth]{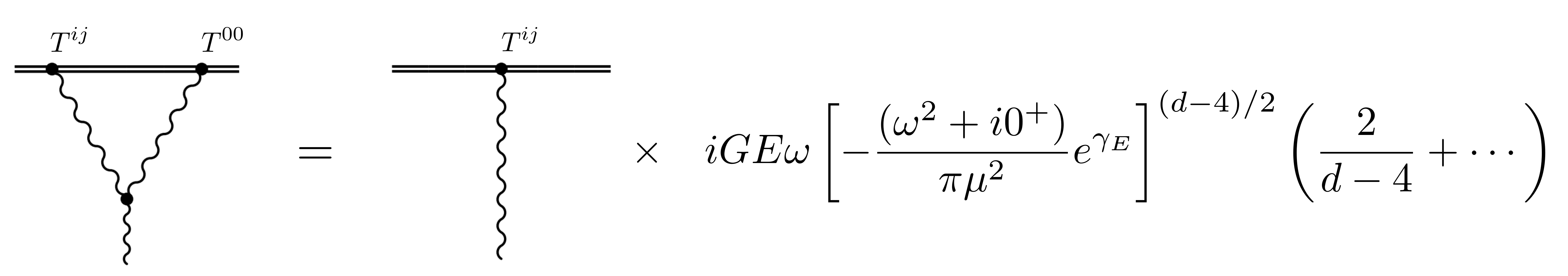} 
      \caption{Universal ultraviolet structure of the hereditary effects (in the far zone) at leading order in the $GE\omega$ expansion. The wavy and (doubly) solid lines represent the gravitational field and binary system, respectively. The latter is described by the (near zone) stress-energy tensor~$T^{\mu\nu}$. Only tail terms contribute to the (ultraviolet) divergence~\cite{tail2}. The ellipses contain finite terms in the $d\to 4$ limit. See \cite{andirad} for more details.} 
      \label{tail1}
            \vspace{-0.4cm}
            \end{center}
\end{figure}

As it was shown some time ago in \cite{andirad}, the tail correction to the GW amplitude becomes
\beq
|{\cal A}_{\rm src}+{\cal A}_{\rm tail}|^2 = (1+2\pi GE\omega ) |{\cal A}_{\rm src}|^2 + {\cal O}\left((GE\omega)^2\right)\label{ampsrc}\,,
\eeq
at leading order in the $(GE\omega)$ expansion in the far zone. The (source) amplitude, ${\cal A}_{\rm src}$, defined by
\beq
i{\cal A}_{\rm src}(\omega,\bk) = -\frac{i}{2} \epsilon^\star_{ij} (\bk) T^{ij}(\omega,\bk)\,,
\eeq
depends upon the near-zone pseudo stress-energy tensor (including also the binding potential modes), $T^{ij}(\omega,\bk)$, as well as the polarization tensor,  $\epsilon_{ij}(\bk)$. The above relation is a consequence of the link depicted in Fig.~\ref{tail1}.\footnote{This is sometimes written in terms of so-called ``radiative" multipole moments associated with the (long-wavelength) expansion of $T^{ij}(\omega,\bk)$ in powers of $\bk$.}\vskip 4pt	

\begin{figure}[t!] 
\begin{center}
\includegraphics[width=0.25\textwidth]{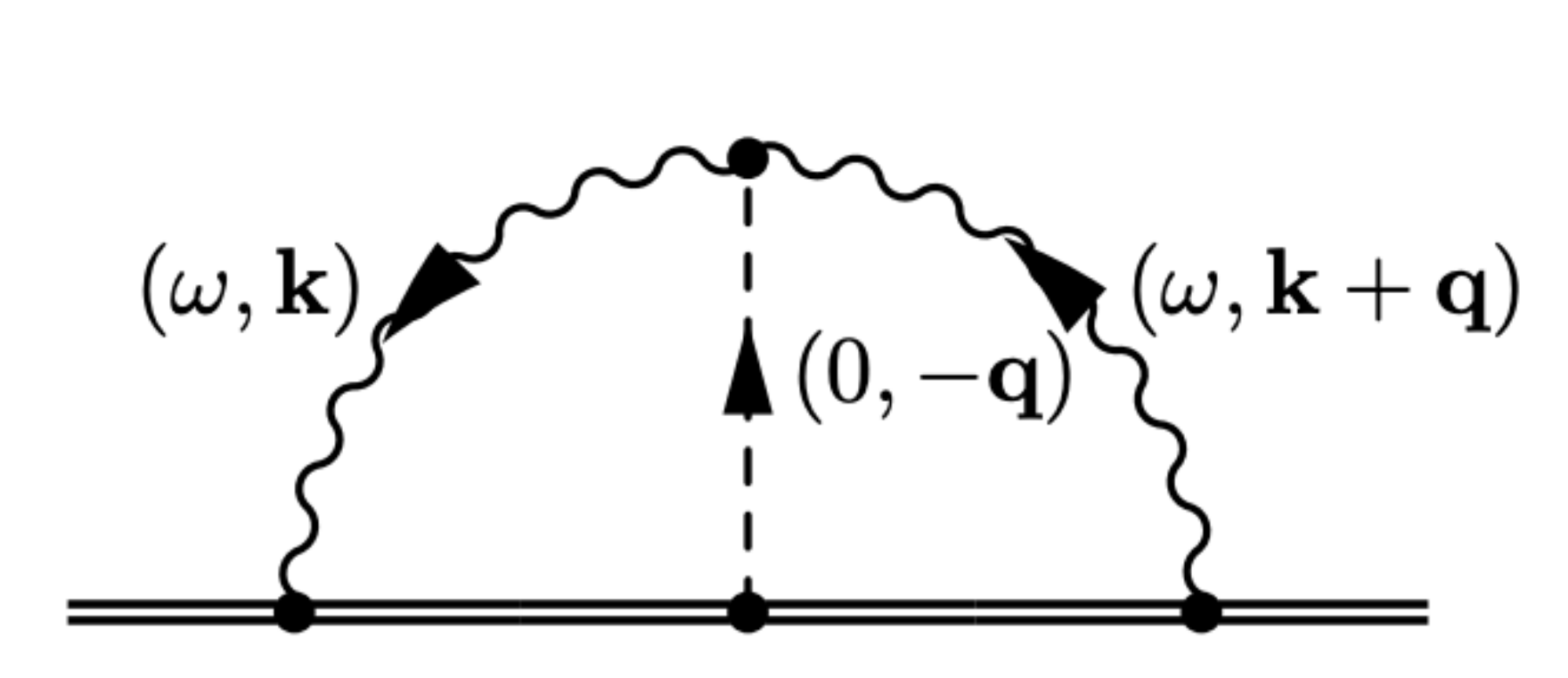} 
      \caption{Correction to the effective action due to the leading tail effect. The wavy line is the (on-shell) radiation mode while the background geometry is sourced by the total energy represented by the dashed line. The causal routing features retarded propagators, see \cite{tail} for more details.} 
      \label{tail2}
            \vspace{-0.4cm}
            \end{center}
\end{figure}

At the same time we have a conservative contribution to the effective action, $S_{\rm eff}^{\rm tail} \equiv - \frac{1}{2\pi} \int_{-\infty}^{+\infty} H_{\rm tail} dt$, with $H_{\rm tail}$ the tail Hamiltonian, shown in Fig.~\ref{tail2}. As discussed in \cite{amps,tail}, the main difference between retarded and Feynman propagators involves the choice of $i0^+$~prescription, yielding a result with a left-over integral over $d\omega$ involving either
\beq
\begin{aligned}
\frac{1}{(d-4)} \left(\sqrt{-(\omega^2+i0^+)}\right)^{(d-4)} &\quad\quad (\rm Feynman)\,,\quad\quad {\rm or}\\
\frac{1}{(d-4)} \left(\sqrt{-(\omega+i0^+)^2}\right)^{(d-4)}&\quad\quad (\rm Retarded)\,,\label{retfey}
\end{aligned}
\eeq 
in dimensional regularization in $d$ dimensions. It is straightforward to show that the difference in the derivation in either case 
only affects the imaginary (dissipative) part. Indeed, the associated factor of $i\pi$ (versus $i\pi\, \text{sign}\,\omega$) is required from the optical theorem, 
\beq
{\rm Im\,\, Fig} \,\ref{tail2} = \frac{1}{2}\int \frac{d\Gamma_{\rm tail}}{d\omega} d\omega \quad\quad (\rm Feynman)\,,
\eeq
in order to match the integrated rate of graviton emission, $\Gamma_{\rm tail}$, due to the tail effect. At the same time, after multiplying by the phase space measure on both sides of \eqref{ampsrc}, we find
\beq
\frac{d\Gamma_{\rm tail}}{d\omega} = 2\pi GE \frac{dE_{\rm src}}{d\omega} + {\cal O}(GE\omega)^2)\,,
\eeq
where $\omega d\Gamma_{\rm src} /d\omega \equiv dE_{\rm src}/d\omega$ is the source energy spectrum. This implies, 
\beq
 {\rm Im\,\, Fig} \, \ref{tail2} = GE \pi \int    \frac{dE_{\rm src}}{d\omega} d\omega  \quad\quad (\rm Feynman) \,.
\eeq
Since the factor of $i\pi$ is intimately connected to the ultraviolet pole (as well as logarithmic correction) in the real part of the effective action, whose residue is unaffected by the choice of propagators, we conclude the leading conservative contribution from the tail effect to the effective action must take on the following universal form:
\beq
{\rm Fig}\, \ref{tail2} = - GE  \int    \frac{dE_{\rm src}}{d\omega} \left( \frac{1}{(d-4)_{\rm UV}} + \log \frac{\omega^2}{\mu^2} - i\pi {\rm sign\, \omega} + \cdots  \right) d\omega \quad\quad (\rm Retarded) \,. \label{retail}
\eeq
Notice that, while we work at leading order in the tail expansion in the radiation region, the expression in \eqref{retail} enters at all PM orders due to ${\cal O}(G_N)$ corrections from the near zone to the source energy flux. From here we conclude
\beq
S^{\rm tail}_{\rm eff, hyp} = - \frac{GE}{(d-4)_{\rm UV}}  \,\Delta E_{\rm hyp} (J,\cE)+\cdots\,,\label{retail2}
\eeq
evaluated on hyperbolic orbits. The above expressions connect the total radiated energy to the ultraviolet pole in the effective action due to tail effects. As we shall see, this universal relation can be used to extend the B2B dictionary to the conservative sector, and vice versa. 

\subsection{Effective to radial action} 

In order to use the relation in \eqref{retail2} we must first connect the value of the effective action to the radial action. This can be achieved by using Hamilton-Jacobi's approach to the motion with the effective action as the generating functional. By~choosing the $\theta=\pi/2$ plane, going to the center-of-mass, and isolating conserved quantities, we have
\beq
S_{\rm eff} = - E \int dt + J \int d\varphi +  2\pi \cS_{r, \rm hyp} \,,
\eeq
where the radial action evaluated on hyperbolic motion is given by
\beq
\cS_{r,\rm hyp} = \frac{2}{2\pi}\int_{\tilde r_-}^\infty p_r dr\,,\label{srhyp}
\eeq
and the factor of $1/2\pi$ added for convenience. Hence, at leading order in the tail expansion,
\beq
\cS^{\rm tail}_{r, \rm hyp} = \frac{1}{2\pi} S_{\rm eff, hyp}^{\rm tail} \,.\label{effr}
\eeq
We can also prove the relationship in \eqref{effr} directly from the definition in \eqref{srhyp}. Let us start by separating the potential and tail contributions to the square of the center-of-mass momentum as
\beq
\bp^2  = p_\infty^2 +  \bp_{\rm pot }^2 (r,\cE)+ \bp_{\rm tail }^2 (r,\cE)\,,
\eeq
such that, to leading order in the tail expansion,
\beq
(\delta p_r)_{\rm tail} = \frac{1}{2}  \frac{\bp_{\rm tail }^2}{p_r} + \cdots \,.
\eeq
We can then compute the correction to the radial action,  
\beq
\cS^{\rm tail}_r = \frac{1}{2\pi} \int_{-\infty}^{+\infty} \dot r dt \, (\delta p_r)_{\rm tail}  = \frac{1}{2\pi}  \int_{-\infty}^{+\infty} dt \left(\partial H_{\rm pot} \over \partial \bp^2\right) \bp_{\rm tail}^2+\cdots\,,
\eeq
where we used \eqref{dotr} and $H_{\rm pot}$ is the Hamiltonian from potential modes. Finally, we must express the momentum in terms of the tail Hamiltonian. Using the relationship \cite{paper1}, 
\beq
\sqrt{\bp^2 - \bp_{\rm pot}^2(r,\cE)- \bp_{\rm tail}^2(r,\cE) + m_1^2} + \sqrt{\bp^2 - \bp_{\rm pot}^2(r,\cE)-\bp_{\rm tail}^2(r,\cE) + m_2^2} = H_{\rm pot} +H_{\rm tail}
\eeq
we have
\beq
H_{\rm tail} = -\left(\partial H_{\rm pot} \over \partial \bp^2\right) \bp_{\rm tail}^2(r,\cE) + \cdots\,, 
\eeq
hence
\beq
\cS_{r,\rm hyp}^{\rm tail} = -\frac{1}{2\pi} \int_{-\infty}^{+\infty}H_{\rm tail} \, dt = \frac{1}{2\pi} S_{\rm eff, hyp}^{\rm tail}\,,
\eeq
at leading order in the tail expansion, yet to all PN orders, as anticipated.\vskip 4pt

The case of elliptic-like motion is a tad more subtle. It is straightforward to arrive at the same condition as in \eqref{retail} for the effective action, however, the connection to the radial action is less direct since the latter involves an integral over a period, whereas the effective action is integrated between minus and plus infinity. Nevertheless, we can follow the exact same steps as above using the tail Hamiltonian integrated over only one period of the bound orbit, thus arriving at a similar relationship as in \eqref{retail2}
\beq
\cS^{\rm tail}_{r, \rm ell} =  \frac{2}{2\pi}\int_{r_-}^{r_+} p_r dr = - \frac{GE}{2\pi(d-4)_{\rm UV}}  \,\Delta E_{\rm ell} (J,\cE)+\cdots\,.\label{retail3}
\eeq

The reader will now immediately realize that the above imply that the pole in the radial/effective action transforms into a similar pole in the tail Hamiltonian, 
\beq
\frac{H_{\rm tail}(r,\bp^2) }{GE} = \frac{{\cal F}_E^{\rm src}(r,\bp^2)}{(d-4)_{\rm UV}} + \cdots\,, \label{hpole}
\eeq
from which we can read off the (averaged) flux for generic orbits.\vskip 4pt Notice the existence of an isotropic gauge for the local-in-time piece of the tail Hamiltonian implies that the flux may be written solely as a function of $r$ and $\bp^2$ (for non-spinning bodies). There is, nonetheless, an important difference with the representation in \S\ref{inverse} regarding the variables used to evaluate the flux. While we use $(r,\cE)$ in \S\ref{inverse}, the derivation from the tail Hamiltonian involves $(r,\bp^2)$. This difference simply results in a mismatch for the coefficients of the PM expansion, which can be easily reconstructed from each other.

\subsection{Local-in-time}
We can now  use the B2B map unveiled in \eqref{b2ben}, and following the manipulations in \eqref{retail2}-\eqref{retail3}, we arrive~at 
\beq
\cS^{\rm tail, pole}_{r, \rm ell}(J,\cE) = \cS^{\rm tail, pole}_{r, \rm hyp}(J,\cE) - \cS^{\rm tail, pole}_{r, \rm hyp}(-J,\cE)\label{b2bloc0}\,.
\eeq
Since the pole contributes to the local-in-time dynamics, we conclude that the latter respects the original B2B dictionary. Furthermore, by performing a multipole expansion of the tail contribution to the effective action and using the B2B link for the radiated energy at each PN order, it is also straightforward to show the B2B correspondence applies to all local-in-time tail effects. Needless to say, the converse is also manifestly true. Following the steps described in \cite{paper1,paper2} for all local-in-time contributions to the PM expansion of the center-of-mass momentum, we find that the B2B map between angle and periastron advance remains valid, hence the map in \eqref{b2bloc0}, which in this case would yield \eqref{b2ben} as a side product.\vskip 4pt

To complete our discussion of local-in-time effects we must also include conservative memory contributions. However, despite its radiative origin, these can be described by the standard Firsov representation, and therefore naturally obeys the original B2B correspondence. We will add a few comments about isolating the conservative part of the memory~in~\S\ref{disc}.\vskip 4pt
As~a~consequence, we find
\beq
\cS^{ \rm loc}_{r, \rm ell}(J,\cE) = \cS^{ \rm loc}_{r, \rm hyp}(J,\cE) - \cS^{\rm loc}_{r, \rm hyp}(-J,\cE)\label{b2bloc}\,.
\eeq
for all local-in-time dynamics, either from potential, tail or memory effects, and generic orbits. \vskip 4pt

Before moving on, let us stress an important point. At first sight the local-in-time dynamics may appear to include also contributions depending on $\log r$. The latter enters in the computation of the conservative dynamics involving potential-only modes \cite{nrgr4pn1,nrgr4pn2}. In~principle, one can readily incorporate these effects in the B2B map of \eqref{b2bloc}. However, this $\log r$ is spurious and is only due to the split into regions.\footnote{The spurious nature is also manifest in the lack of factors of $\log J$ (or $\log b$'s) in the total unbound radial action. After noticing that the $\log J$'s are directly linked to the poles arising in the $d\to 4$ limit, their disappearance is ultimately rooted in the same cancelation between infrared/ultraviolet divergences.}  There are, nonetheless, crucial logarithmic corrections to the radial action, but involving instead the binding energy (see below). Because of these reasons, it has become customary in the PN literature, e.g. \cite{binidam1,binidam2}, to define the non-local-in-time contribution to the effective action from the (leading) tail as follows
\beq
 S^{\rm nloc}_{\rm eff} \equiv - GE \int \frac{dE_{\rm src}}{d\omega} \log (4r^2\omega^2e^{\gamma_E})d\omega\,,\label{snloc}
\eeq
mixing the logarithmic terms from both potential and radiation modes (The factor of $e^{\gamma_E}$ and $\log 2$ are  follow the PN conventions.) In this fashion, all the other pieces are collected as local-in-time contributions which are thus devoid of any logarithmic correction. As we have shown in \eqref{retail}, the expression in \eqref{snloc} retains its universal character in a PM scheme and therefore it can be generally adopted to describe non-local-in-time effects. We will use this definition in what follows. 
\subsection{Large-eccentricity limit}

 At this point the reader may wonder whether the B2B dictionary can remain valid also for the non-local-in-time dynamics induced by \eqref{snloc}. As we demonstrate in what follows, that is indeed the case. However, there is a major caveat. The B2B map only holds in the limit of large angular momentum. In order to demonstrate this, we start by re-writing the expression in \eqref{snloc} in terms of a non-local Hamiltonian, 
\beq
\cS_{r,\rm hyp}^{\rm nloc}(J,\cE) = -\frac{2}{2\pi} \int_{\tilde r_-}^{\infty} H_{\rm nloc}(r,J,\cE) \frac{dr}{\dot r}\,.
\eeq
In light of our previous manipulations with the fluxes in \S\ref{sec:rad}, and depending on the given  $J$-parity of  $H_{\rm nloc}$, it is clear a B2B-type correspondence must apply. Indeed, we find 
\beq
H_{\rm nloc}(r,J,\cE) = + H_{\rm nloc}(r,-J,\cE)\,,\label{hnloc}
\eeq
in the large $J$ expansion, such that in this particular limit the relation in \eqref{b2bloc} remains the same for non-local-in-time terms, which implies for the full radial action
\beq
\cS^{ \rm full}_{r, \rm ell}(J,\cE) = \cS^{ \rm full}_{r, \rm hyp}(J,\cE) - \cS^{\rm full}_{r, \rm hyp}(-J,\cE) \label{b2bnloc}\,.
\eeq
The proof of \eqref{hnloc} proceeds as follows. As discussed in e.g. \cite{Damour:2015isa,binidam1,binidam2}, after truncating the derivation of the non-local (conservative) Hamiltonian to a given order in the PN/PM expansion, one can then construct an {\it effectively local} counter-part describing the evolution of the system to the same PN/PM order.  Once the effective Hamiltonian evaluated for unbound motion is known, we can then construct an isotropic gauge in which the latter becomes a function of $r$ and $\bp^2$ only, which then fulfills the condition in \eqref{hnloc}. (The condition in \eqref{hnloc} is also satisfied  for the gauges in which $(H_{\rm nloc})_{\rm eff}$ is given as a function of $r,\bp^2,p_r^2$.)

\subsection{Logarithms}\label{seclog}

Unlike local effects, non-local tail interactions cause the B2B radial action in the large-eccentricity limit fail to describe generic bound orbits. In particular, non-local terms yield different values for hyperbolic-like and (the more phenomenologically relevant case of) quasi-circular motion. However, somewhat remarkably, the large-eccentricity limit does captures certain non-local contributions for generic orbits, even those with small eccentricities, notably those involving the (leading) logarithms in the binding energy. These terms inherit the universal structure in \eqref{retail2} and \eqref{retail3}, yielding (with $v_\infty \equiv \sqrt{\gamma^2-1}\,$)
\beq
\frac{\cS_{r, \rm hyp/ell}^{\log}}{GM^2\nu} = -\frac{\Gamma}{\pi\nu} \frac{\Delta E_{\rm hyp/ell}}{M} \log |v_\infty|\,,\label{logsr}
\eeq
which is the left over after the cancelation of logarithms in the angular momentum accompanying the IR/UV poles. They can also be shown to appear directly from the scaling with the velocity of radiation modes, see~\cite{4pmeft,4pmtail}. Using the map between total radiated energy in \eqref{b2ben} the B2B relation in \eqref{b2bloc} is thus manifest. As a result, not only all of the local-in-time effects may be readily mapped between unbound and bound motion for generic orbits, also the trademark logarithmic terms. Furthermore, the latter are related by the same B2B formulae derived in \cite{paper1,paper2}.

\section{Post-Newtonian/Minkowskian}\label{secpnm}

We provide here extensive evidence for the validity of the extended B2B map within the realm of the PN and PM expansions. Let us remind the reader of the definitions 
\beq
\epsilon \equiv -2\cE = -2\left(\frac{E-M}{M\nu}\right)\,, \quad\quad j \equiv \frac{J}{GM^2\nu}\,,  \quad\quad \ell \equiv \frac{L}{GM^2\nu}\,, \quad \quad \tilde a_\pm \equiv \frac{1}{GM} (a_1 \pm a_2)\,,
\eeq
for the reduced binding energy and total, orbital and spin (canonical) angular momentum, respectively. Finite size effects due to spin are parameterized in terms of the $\kappa_\pm$ parameters introduced in \cite{pmefts}.  For the sake of notation, we do not distinguish here between $e$ and $\tilde e$ for hyperbolic- and elliptic-like motion. Moreover, for convenience, some of the terms below are written in terms of the {\it Newtonian} eccentricity $e_N^2 \equiv 1+ 2\cE j^2$ \cite{cho3pn}. (Since $e = e_N + \cdots$, this is simply a reshuffling of various coefficients in the $1/j$ expansion.) To emphasize the PN orders we will use the following color coding: 0PN, \textcolor{1PN}{1PN}, \textcolor{2PN}{2PN}, \textcolor{3PN}{3PN}, \textcolor{4PN}{4PN}, \textcolor{5PN}{5PN}, \textcolor{6PN}{6PN}.

\subsection{Radiated energy}

The spin-independent radiated energy due to the source multipole moments (without including tail terms) has been computed to \textcolor{3PN}{3PN}, both for hyperbolic \cite{cho3pn} and elliptic \cite{Arun:2007sg}. The results can be written as a function of $e_N,j,$ and $\cE$ as follows
\begingroup
\allowdisplaybreaks
\small
\begin{align}&\Delta E_\text{hyp}(j,\cE)
=\frac{M\,\nu^2}{15}\,\Bigg[\frac{850 \sqrt{2} \sqrt{\mathcal{E}}}{j^6}+\frac{2692 \sqrt{2} \mathcal{E}^{3/2}}{3 j^4}+\left(\frac{850}{j^7}+\frac{1464 \mathcal{E}}{j^5}+\frac{296 \mathcal{E}^2}{j^3}\right)
   \cos ^{-1}\left(-\frac{1}{e_N}\right)\\
   &+\textcolor{1PN}{ \,\frac{\sqrt{2} \mathcal{E}^{5/2} (2506431-3009160 \nu )}{105 (1+2 \mathcal{E} j^2) j^4}+\frac{\mathcal{E}^{3/2} (182337-140480 \nu )}{3 \sqrt{2}
   (1+2 \mathcal{E} j^2) j^6}-\frac{7 \sqrt{\mathcal{E}} (-5763+3220 \nu )}{2 \sqrt{2} (1+2 \mathcal{E} j^2) j^8}}\notag\\
   &\textcolor{1PN}{-\frac{2 \sqrt{2} \mathcal{E}^{7/2} (-89907+156380 \nu )}{35 (1+2 \mathcal{E} j^2) j^2}+\Bigg(\frac{\mathcal{E}
   \left(\frac{33885}{2}-15900 \nu \right)}{j^7}+\frac{\mathcal{E}^2 \left(\frac{46617}{7}-10464 \nu \right)}{j^5}}\notag\\
   &\textcolor{1PN}{+\frac{\frac{40341}{4}-5635 \nu }{j^9}+\frac{\mathcal{E}^3
   \left(\frac{4786}{7}-888 \nu \right)}{j^3}\Bigg) \cos ^{-1}\left(-\frac{1}{e_N}\right)}\notag\\
   &\textcolor{2PN}{+\frac{2 \sqrt{2} \mathcal{E}^{11/2} \left(2039036-4763297 \nu
   +6219570 \nu ^2\right)}{245 (1+2 \mathcal{E} j^2)^2}
   +\frac{\sqrt{\mathcal{E}} \left(29198255-32514426 \nu +6906060 \nu ^2\right)}{168 \sqrt{2} (1+2 \mathcal{E} j^2)^2 j^{10}}}\notag\\
   &\textcolor{2PN}{+\frac{\mathcal{E}^{3/2}
   \left(605244551-845377092 \nu +233615340 \nu ^2\right)}{756 \sqrt{2} (1+2 \mathcal{E} j^2)^2 j^8}+\frac{\mathcal{E}^{5/2} \left(6449654885-13029505497 \nu +4951304820 \nu ^2\right)}{5670
   \sqrt{2} (1+2 \mathcal{E} j^2)^2 j^6}}\notag\\
   &\textcolor{2PN}{+\frac{\sqrt{2} \mathcal{E}^{9/2} \left(208988912-6547442643 \nu +6196124970 \nu ^2\right)}{19845 (1+2 \mathcal{E} j^2)^2 j^2}}\textcolor{2PN}{+\frac{\mathcal{E}^{7/2}
   \left(9310241671-39312545895 \nu +22349042100 \nu ^2\right)}{19845 \sqrt{2} (1+2 \mathcal{E} j^2)^2 j^4}}\notag\\
   &\textcolor{2PN}{+\Bigg(\frac{5 \mathcal{E}^4}{21
   j^3} \left(8344-7179 \nu +7770 \nu ^2\right)+\frac{1}{j^{11}}\left(\frac{29198255}{336}-\frac{774153 \nu }{8}+\frac{82215 \nu ^2}{4}\right)}\notag\\
   &\textcolor{2PN}{+\frac{\mathcal{E}^3 }{j^5}\left(\frac{59900}{21}-\frac{219314 \nu }{7}+38412 \nu
   ^2\right)+\frac{\mathcal{E}}{j^9} \left(\frac{11947909}{108}-\frac{2838577 \nu }{12}+85995 \nu ^2\right)}\notag\\
   &\textcolor{2PN}{+\frac{5 \mathcal{E}^2}{252 j^7} \left(736055-7764219 \nu +5348700
   \nu ^2\right)\Bigg) \cos ^{-1}\left(-\frac{1}{e_N}\right)}\notag\\
  &\textcolor{3PN}{+ \frac{\mathcal{E}^{9/2} }{2910600 \sqrt{2} (1+2 \mathcal{E} j^2)^3 j^4}\Big(388429282846397-2640 \left(58245623491+813280797 \pi
   ^2\right) \nu}\notag\\
   &\textcolor{3PN}{ +161210743695540 \nu ^2-51997288820400 \nu ^3\Big)+\frac{\mathcal{E}^{11/2}}{1455300 \sqrt{2} (1+2 \mathcal{E} j^2)^3 j^2} \big(84259836657351}\notag\\
   &\textcolor{3PN}{-220
   \left(7446870460+5591391687 \pi ^2\right) \nu +36932569900380 \nu ^2-18771891415200 \nu ^3\big)}\notag\\
   &\textcolor{3PN}{+\frac{\mathcal{E}^{13/2}}{2182950 \sqrt{2}
   (1+2 \mathcal{E} j^2)^3}
   \big(20497310013607-220 \left(-32611235596+2325312171 \pi ^2\right) \nu}\notag\\
   & \textcolor{3PN}{+10835961651900 \nu ^2-9517996719000 \nu ^3\big)+\frac{\mathcal{E}^{7/2}}{317520
   \sqrt{2} (1+2 \mathcal{E} j^2)^3 j^6} \big(49134463233397}\notag\\
   &\textcolor{3PN}{+8 \left(-4148934390079+7132373325 \pi ^2\right) \nu +18910687025664 \nu ^2-4244390811360 \nu ^3\big)}\notag\\
   &\textcolor{3PN}{+\frac{\mathcal{E}^{5/2} }{151200 \sqrt{2} (1+2 \mathcal{E} j^2)^3 j^8}\Big(14498840196199+20 \left(-622188698008+4083668757 \pi ^2\right) \nu }\notag\\
   &\textcolor{3PN}{+5075295687000 \nu ^2-844365060000 \nu
   ^3\Big)+\frac{\mathcal{E}^{3/2} }{302400 \sqrt{2} (1+2 \mathcal{E} j^2)^3 j^{10}}\big(9133770967709}\notag\\
   &\textcolor{3PN}{+1100 \left(-8061884522+70585641 \pi ^2\right) \nu +2888074488600 \nu ^2-371614824000
   \nu ^3\big)}\notag\\
   &\textcolor{3PN}{+\frac{\mathcal{E}^{15/2} j^2}{24255 \sqrt{2}
   (1+2 \mathcal{E} j^2)^3} \left(744936393-5027892760 \nu +9355109148 \nu ^2-10875359880 \nu ^3\right)}\notag\\
   &\textcolor{3PN}{+\frac{\sqrt{\mathcal{E}} \left(153014201249+440 \left(-361091813+3587787 \pi ^2\right) \nu +43750060200 \nu ^2-4482324000 \nu ^3\right)}{40320 \sqrt{2} (1+2 \mathcal{E} j^2)^3
   j^{12}}}\notag\\
   &\textcolor{3PN}{+\Big(\frac{\mathcal{E}^2}{j^9} \left(\frac{178442872459}{30240}-\frac{5}{432} \left(142791046+2595915 \pi ^2\right) \nu +\frac{23677969 \nu ^2}{12}-607110 \nu
   ^3\right)}\notag\\
   &\textcolor{3PN}{+\frac{\mathcal{E}^3}{j^7} \left(\frac{10364987867}{5040}+\left(\frac{119338465}{378}-\frac{259735 \pi ^2}{8}\right) \nu +\frac{19930745 \nu ^2}{28}-444125
   \nu ^3\right)}\notag\\
   &\textcolor{3PN}{+\frac{\mathcal{E}}{j^{11}} \left(\frac{137076707247}{22400}+\frac{11}{224} \left(-84607982+488187 \pi ^2\right) \nu +\frac{30107727 \nu ^2}{16}-\frac{635985
   \nu ^3}{2}\right)}\notag\\
   &\textcolor{3PN}{+\frac{\mathcal{E}^4 }{j^5}\left(\frac{11296885277}{92400}+\left(\frac{534595}{14}-\frac{12177 \pi ^2}{4}\right) \nu +\frac{2595759 \nu
   ^2}{28}-103395 \nu ^3\right)}\notag\\
   &\textcolor{3PN}{+\frac{1}{j^{13}}\,\big(\frac{25574348567}{11520}+\frac{11 \left(-361091813+3587787 \pi ^2\right) \nu }{2016}+\frac{17361135 \nu
   ^2}{32}-\frac{444675 \nu ^3}{8}\big)}\notag\\
   &\textcolor{3PN}{+\frac{\mathcal{E}^5}{j^3}\left(\frac{2075735}{1848}-\frac{11920 \nu }{3}+\frac{45467 \nu ^2}{14}-3256 \nu ^3\right)\Big)
   \cos ^{-1}\left(-\frac{1}{e_N}\right)}\notag\\
   &\textcolor{3PN}{+\left(\frac{161249 \sqrt{2} \sqrt{\mathcal{E}}}{j^{12}}+\frac{7455118 \sqrt{2} \mathcal{E}^{3/2}}{15 j^{10}}+\frac{6414436 \sqrt{2}
   \mathcal{E}^{5/2}}{15 j^8}+\frac{140201672 \sqrt{2} \mathcal{E}^{7/2}}{1575 j^6}\right) \log \left(\frac{2 \mathcal{E}}{e_N\,}\right)}\notag\\
   &\textcolor{3PN}{+\left(-\frac{161249}{j^{13}}-\frac{3022536
   \mathcal{E}}{5 j^{11}}-\frac{2104904 \mathcal{E}^2}{3 j^9}-\frac{5406496 \mathcal{E}^3}{21 j^7}-\frac{508464 \mathcal{E}^4}{35 j^5}\right)} \notag\\
   &\textcolor{3PN}{\times\left(\text{Cl}_2\left(2 \cos
   ^{-1}\left(-\frac{1}{e_N}\right)\right)+\cos ^{-1}\left(-\frac{1}{e_N}\right) \log \left(\frac{2(e_N^2-1)\, }{e_N\,\mathcal{E}}\right)\right)}\Bigg]\,.\nn
\end{align}
\endgroup
The Clausen function of order 2 is defined by the integral representation,
\begin{align}
\text{Cl}_2(z) \equiv -\int^z_0\,dy\,\log\left|2\,\sin\frac{y}{2}\right|\,.
\end{align}
On the other hand, for a period of an elliptic orbit we have
\begingroup
\allowdisplaybreaks
\small
\begin{align}
&\Delta E_\text{ell}(j,\cE)=\frac{M\,\nu^2}{15}\,\Bigg[\frac{850 \pi }{j^7}+\frac{1464 \mathcal{E} \pi }{j^5}+\frac{296 \mathcal{E}^2 \pi }{j^3}+\textcolor{1PN}{ \frac{\mathcal{E}^2 \pi }{j^5} \left(\frac{46617}{7}-10464 \nu \right)}\\
&\textcolor{1PN}{+\frac{7
   \pi  (5763-3220 \nu )}{4 j^9}+\frac{15 \mathcal{E} \pi  (2259-2120 \nu )}{2 j^7}+\frac{\mathcal{E}^3 \pi }{j^3} \left(\frac{4786}{7}-888 \nu \right)}\notag\\
   &\textcolor{2PN}{+
 \frac{5 \mathcal{E}^4 \pi  \left(8344-7179 \nu +7770 \nu ^2\right)}{21 j^3}+\frac{2 \mathcal{E}^3 \pi  \left(29950-328971 \nu +403326 \nu ^2\right)}{21
   j^5}}\notag\\
   &\textcolor{2PN}{+\frac{5 \mathcal{E}^2 \pi  \left(736055-7764219 \nu +5348700 \nu ^2\right)}{252 j^7}+\frac{\pi  \left(29198255-32514426 \nu +6906060 \nu ^2\right)}{336
   j^{11}}}\notag\\
   &\textcolor{2PN}{+\frac{\mathcal{E} \pi  \left(11947909-25547193 \nu +9287460 \nu ^2\right)}{108 j^9}}\notag\\
   &\textcolor{3PN}{ -\frac{161249 \sqrt{2} \sqrt{-\mathcal{E}} \pi
   }{j^{12}}-\frac{7455118 \sqrt{2} \sqrt{-\mathcal{E}} \mathcal{E} \pi }{15 j^{10}}-\frac{6414436 \sqrt{2} \sqrt{-\mathcal{E}} \mathcal{E}^2 \pi }{15 j^8}}\notag\\
   &\textcolor{3PN}{-\frac{140201672
   \sqrt{2} \sqrt{-\mathcal{E}} \mathcal{E}^3 \pi }{1575 j^6}+\frac{\mathcal{E} \pi}{22400 j^{11}}  \big(137076707247+1100 \left(-84607982+488187 \pi ^2\right) \nu}\notag\\
   &\textcolor{3PN}{ +42150817800 \nu
   ^2-7123032000 \nu ^3\big)+\frac{\pi}{80640 j^{13}}\,\big(179020439969}\notag\\
   &\textcolor{3PN}{+440 \left(-361091813+3587787 \pi ^2\right) \nu +43750060200 \nu ^2-4482324000 \nu
   ^3\big)}\notag\\
   &\textcolor{3PN}{+\frac{\mathcal{E}^5 \pi}{j^3} \left(\frac{2075735}{1848}-\frac{11920 \nu }{3}+\frac{45467 \nu ^2}{14}-3256 \nu ^3\right)}\notag\\
   &\textcolor{3PN}{+\frac{\mathcal{E}^2}{j^9}
   \left(-\frac{1}{16} \left(480725 \pi ^3 \nu \right)+\pi  \left(\frac{178442872459}{30240}-\frac{356977615 \nu }{216}+\frac{23677969 \nu ^2}{12}-607110 \nu
   ^3\right)\right)}\notag\\
   &\textcolor{3PN}{+\frac{\mathcal{E}^3 }{j^7}\left(-\frac{1}{8} \left(259735 \pi ^3 \nu \right)+\pi  \left(\frac{10364987867}{5040}+\frac{119338465 \nu
   }{378}+\frac{19930745 \nu ^2}{28}-444125 \nu ^3\right)\right)}\notag\\
   &\textcolor{3PN}{+\frac{\mathcal{E}^4 }{j^5}\left(-\frac{1}{4} \left(12177 \pi ^3 \nu \right)+\pi 
   \left(\frac{11296885277}{92400}+\frac{534595 \nu }{14}+\frac{2595759 \nu ^2}{28}-103395 \nu ^3\right)\right)}\notag\\
   &\textcolor{3PN}{+\left(\frac{161249 \pi }{j^{13}}+\frac{3022536
   \mathcal{E} \pi }{5 j^{11}}+\frac{2104904 \mathcal{E}^2 \pi }{3 j^9}+\frac{5406496 \mathcal{E}^3 \pi }{21 j^7}+\frac{508464 \mathcal{E}^4 \pi }{35 j^5}\right) \times \log
   \left(\frac{(1+\sqrt{1-e_N^2})\mathcal{E}}{2 (e_N^2-1)}\right)}\Bigg]\,.\nn
\end{align}
\endgroup
In order to apply the B2B correspondence in \eqref{b2ben} we must exercise some care with the analytic continuation in the angular momentum and binding energy. In particular, notice that we have expanded the polynomial factors in $1/j$ explicitly, while keeping the special functions in terms of the eccentricity. For the latter we apply the analytic continuation according to the B2B dictionary using the definition of the principal values, e.g.
\beq 
\cos^{-1}(z)=-i\,\log \left(z+i\sqrt{1-z^2}\right)\,,
\eeq
such that
\beq
\label{arcos}
\begin{aligned}
\cos^{-1}\left(\frac{1}{e}\right)+\cos^{-1}\left(-\frac{1}{e}\right)=\pi\,,\quad
\log\left(\frac{1}{e\,}\right)-\log\left(-\frac{1}{e\,}\right)=i\,\pi\,,
\end{aligned}
\eeq
gives the relation
\begin{align}
&\quad\quad\quad\cos^{-1}\!\left(-\frac{1}{e}\right) \log \left(\frac{2(e^2-1)\, }{e\,\mathcal{E}}\right)+\cos^{-1}\left(\frac{1}{e}\right) \log \left(-\frac{2(e^2-1)\, }{e\,\mathcal{E}}\right)\nn\\
&=\cos^{-1}\!\left(-\frac{1}{e}\right)\,\left(\log \left(\frac{2(e^2-1)\, }{e\,\mathcal{E}}\right)- \log \left(-\frac{2(e^2-1)\, }{e\,\mathcal{E}}\right)\right)+\pi\,\log \left(-\frac{2(e^2-1)\, }{e\,\mathcal{E}}\right)\nn\\
&=\cos^{-1}\!\left(-\frac{1}{e}\right)\,i\,\pi+\pi\, \log \left(-\frac{2(e^2-1)\, }{e\,\mathcal{E}}\right)=\pi \log \left(-\frac{1}{e}+i\sqrt{1-\frac{1}{e^2}}\right)+\pi\, \log \left(-\frac{2(e^2-1)\, }{e\,\mathcal{E}}\right)\nn\\
&=\pi\,\log \left(-2\frac{(1-\sqrt{1-e^2})(1-e^2)}{e^2 \mathcal{E}}\right)=-\pi\,\log \left(\frac{(1+\sqrt{1-e^2})\mathcal{E}}{2(e^2-1)}\right)\,,
\end{align}
which links hyperbolic and elliptic orbits. Another important relation is given by 
\begin{align}
\text{Cl}_2 \left(2\,\cos^{-1}{\left(\frac{-1}{e}\right)}\right)+\text{Cl}_2 \left(2\,\cos^{-1}{\left(\frac{1}{e}\right)}\right)=0\,,
\end{align}
which altogether disappears from the elliptic case. This is proved by noticing 
\begin{align}
\text{Cl}_2 \left(2\,\cos^{-1}{\left(\frac{-1}{e}\right)}\right)&=\text{Cl}_2 \left(2\,\pi-2\,\cos^{-1}{\left(\frac{1}{e}\right)}\right)\nn\\
&=-\int^{2\,\pi-2\,\cos^{-1}{\left(\frac{1}{e}\right)}}_0 dz\,\log\left|2\sin\frac{z}{2}\right|
=-\int^{-2\,\cos^{-1}{\left(\frac{1}{e}\right)}}_{-2\pi} dz\,\log\left|2\sin\frac{z+2\pi}{2}\right|\nn\\
&=-\int^{2\,\pi}_0 dz\,\log\left|2\sin\frac{z}{2}\right|-\int^{-2\,\cos^{-1}{\left(\frac{1}{e}\right)}}_0 dz\,\log\left|2\sin\frac{2\pi+z}{2}\right|\nn\\
&=-\int^{-2\,\cos^{-1}{\left(\frac{1}{e}\right)}}_0 dz\,\log\left|2\sin\frac{z}{2}\right|=\int^{2\,\cos^{-1}{\left(\frac{1}{e}\right)}}_0 dz\,\log\left|2\sin\frac{z}{2}\right|\nn\\
&=-\text{Cl}_2 \left(2\,\cos^{-1}{\left(\frac{1}{e}\right)}\right)
\end{align}
where we used the identity $\int^{2\,\pi}_0 dz\,\log|2\sin\frac{z}{2}|=0$.\vskip 4pt Other terms, such as those involving~$\sqrt{\cE}$, produce imaginary pieces for bound orbits which are either canceled against similar complex terms or by the B2B relations. After all of these manipulations, we find that the B2B correspondence in \eqref{b2ben} is neatly fulfilled to \textcolor{3PN}{3PN} order. A similar check was discussed in \cite{parra2} in the context of the PM expansion. We confirm the results in \cite{parra2} are consistent with the PN values presented here.

\subsection{Radiated angular momentum}

The same logic applies to the angular momentum. For non-spinning bodies we find 
\begingroup
\allowdisplaybreaks
\small
\begin{align}
&\Delta J_\text{hyp}(j,\cE)=\frac{8\,G\,M^2\,\nu^2}{5}\,\Bigg[\frac{15 \sqrt{2} \sqrt{\mathcal{E}}}{j^3}+\frac{4 \sqrt{2} \mathcal{E}^{3/2}}{j}+\left(\frac{15}{j^4}+\frac{14 \mathcal{E}}{j^2}\right) \cos ^{-1}\left(-\frac{1}{e_N}\right)\\
&\textcolor{1PN}{+
  \frac{\mathcal{E}^{3/2} \left(5541 \sqrt{2}-8620 \sqrt{2} \nu \right)}{72 j^3}+\frac{\sqrt{\mathcal{E}} \left(4953 \sqrt{2}-4748 \sqrt{2} \nu \right)}{48
   j^5}+\frac{\mathcal{E}^{5/2} \left(109 \sqrt{2}-35 \sqrt{2} \nu \right)}{7 j}}\notag\\
   &\textcolor{1PN}{+\frac{\sqrt{\mathcal{E}} \left(9 \sqrt{2}+\sqrt{2} \nu \right)}{(1+2 \mathcal{E} j^2)
   j^5}+\left(\frac{\mathcal{E}^2 (4283-3976 \nu )}{84 j^2}+\frac{\mathcal{E} (535-748 \nu )}{4 j^4}-\frac{5 (-1077+940 \nu )}{48 j^{6}}\right)\cos^{-1}\left(-\frac{1}{e_N}\right)}\notag\\
   &\textcolor{2PN}{+ \frac{\sqrt{\mathcal{E}} \left(123507 \sqrt{2}-23678 \sqrt{2} \nu -2289 \sqrt{2} \nu ^2\right)}{1344 (1+2 \mathcal{E} j^2)
   j^{7}}+\frac{\sqrt{\mathcal{E}} \left(-81 \sqrt{2}-18 \sqrt{2} \nu -\sqrt{2} \nu ^2\right)}{8 (1+2 \mathcal{E} j^2)^2 j^{7}}}\notag\\
   &\textcolor{2PN}{+\frac{\mathcal{E}^{7/2} \left(3245 \sqrt{2}-5886 \sqrt{2} \nu
   +3213 \sqrt{2} \nu ^2\right)}{504 j}+\frac{\mathcal{E}^{5/2} \left(-439377 \sqrt{2}-1828552 \sqrt{2} \nu +2246335 \sqrt{2} \nu ^2\right)}{5040
   j^3}\notag}\\
   &\textcolor{2PN}{+\frac{\sqrt{\mathcal{E}} \left(15340289 \sqrt{2}-38231694 \sqrt{2} \nu +14143059 \sqrt{2} \nu ^2\right)}{36288 j^{7}}}\notag\\
   &\textcolor{2PN}{+\frac{\mathcal{E}^{3/2} \left(-12397361
   \sqrt{2}-59811138 \sqrt{2} \nu +53024895 \sqrt{2} \nu ^2\right)}{54432 j^5}}\notag\\
   &\textcolor{2PN}{+\Big(\frac{\mathcal{E}^3 \left(5308-32877 \nu +28308 \nu ^2\right)}{252 j^2}+\frac{5
   \mathcal{E} \left(-2999-152946 \nu +106848 \nu ^2\right)}{432 j^{6}}}\notag\\
   &\textcolor{2PN}{+\frac{1307683-2782332 \nu +1005480 \nu ^2}{2592 j^{8}}+\frac{\mathcal{E}^2 \left(-404980-1181889
   \nu +1438668 \nu ^2\right)}{1512 j^4}\Big) \cos ^{-1}\left(-\frac{1}{e_N}\right)}\notag\\
   &\textcolor{3PN}{+\left(\frac{3531 \sqrt{\mathcal{E}}}{\sqrt{2} j^{9}}+\frac{10700
   \sqrt{2} \mathcal{E}^{3/2}}{3 j^{7}}+\frac{421366 \sqrt{2} \mathcal{E}^{5/2}}{315 j^5}\right)\,\log
   \left(\frac{2 \,\mathcal{E}}{e_N}\right)}\notag\\
   &\textcolor{3PN}{-\left(\frac{3531}{2 j^{10}}+\frac{14231 \mathcal{E}}{3
   j^{8}}+\frac{21614 \mathcal{E}^2}{7 j^{6}}+\frac{9844 \mathcal{E}^3}{35 j^4}\right) \bigg(\log \left(\frac{2(e_N^2-1) }{e_N\,\mathcal{E}}\right) \cos ^{-1}\left(-\frac{1}{e_N}\right)+\text{Cl}_2\left(2 \cos
   ^{-1}\left(-\frac{1}{e_N}\right)\right)\bigg)}\notag\\
   &\textcolor{3PN}{+\frac{\mathcal{E}^{5/2}}{2177280 j^5} \bigg(10005567679 \sqrt{2}+(6156118418 \sqrt{2}
   -202155912 \sqrt{2} \pi ^2) \nu +12209691726 \sqrt{2} \nu ^2}\notag\\
   &\textcolor{3PN}{-10620524250 \sqrt{2} \nu ^3\bigg)+\frac{\mathcal{E}^{3/2}}{1451520
   j^{7}} \bigg(24486609623
   \sqrt{2}+(-18545026170 \sqrt{2} +252462420 \sqrt{2} \pi ^2 )\nu }\notag\\
   &\textcolor{3PN}{+18259461090 \sqrt{2} \nu ^2-6662875590 \sqrt{2} \nu ^3\bigg)+\frac{\sqrt{\mathcal{E}} }{2903040 j^{9}}\big(37783086543 \sqrt{2}}\notag\\
   &\textcolor{3PN}{+(-52725330410 \sqrt{2}  +834567300 \sqrt{2} \pi ^2 )\nu +19852063650 \sqrt{2} \nu ^2-3342659670 \sqrt{2}
   \nu ^3\big)}\notag\\
   &\textcolor{3PN}{+\frac{\mathcal{E}^{7/2} }{282240 j^3}\left(-106774701 \sqrt{2}+33275322 \sqrt{2} \nu +305926390 \sqrt{2} \nu ^2-341077730 \sqrt{2} \nu
   ^3\right)}\notag\\
   &\textcolor{3PN}{+\frac{\mathcal{E}^{9/2} \left(-6973 \sqrt{2}-35695 \sqrt{2} \nu +334521 \sqrt{2} \nu ^2-172557 \sqrt{2} \nu ^3\right)}{22176
   j}}\notag\\
   &\textcolor{3PN}{+\frac{\sqrt{\mathcal{E}} \left(244427427 \sqrt{2}-165077725 \sqrt{2} \nu +1487808 \sqrt{2} \pi ^2 \nu -6469083 \sqrt{2} \nu ^2-134379 \sqrt{2} \nu
   ^3\right)}{290304 (1+2 \mathcal{E} j^2) j^{9}}}\notag\\
   &\textcolor{3PN}{+\frac{\sqrt{\mathcal{E}} \left(729 \sqrt{2}+243 \sqrt{2} \nu +27 \sqrt{2} \nu ^2+\sqrt{2} \nu ^3\right)}{48 (1+2 \mathcal{E} j^2)^3
   j^{9}}}\notag\\
   &\textcolor{3PN}{+\frac{\sqrt{\mathcal{E}} \left(-772659 \sqrt{2}+48105 \sqrt{2} \nu +22507 \sqrt{2} \nu ^2+847 \sqrt{2} \nu ^3\right)}{5376 (1+2 \mathcal{E} j^2)^2 j^{9}}}\notag\\
   &\textcolor{3PN}{+\Bigg(\frac{\mathcal{E}^2
   \left(2082786797-44373700 \nu -8136450 \pi ^2 \nu +1398279600 \nu ^2-910576800 \nu ^3\right)}{120960 j^{6}}}\notag\\
   &\textcolor{3PN}{+\frac{\mathcal{E}^3 \left(90979951+255064600 \nu
   -5811750 \pi ^2 \nu +433045800 \nu ^2-492912000 \nu ^3\right)}{151200 j^4}}\notag\\
   &\textcolor{3PN}{+\frac{5567205457-6037270480 \nu +94382820 \pi ^2 \nu +2200128840 \nu ^2-371498400 \nu
   ^3}{322560 j^{10}}}\notag\\
   &\textcolor{3PN}{+\frac{\mathcal{E} \left(1737906083-1253646560 \nu +18597600 \pi ^2 \nu +889119000 \nu ^2-277754400 \nu ^3\right)}{51840 j^{8}}}\notag\\
   &\textcolor{3PN}{+\frac{\mathcal{E}^4
   \left(-227005-5072672 \nu +11771892 \nu ^2-9953328 \nu ^3\right)}{44352 j^2}\Bigg) \cos ^{-1}\left(-\frac{1}{e_N}\right)}\Bigg]\,,\notag
\end{align}
\endgroup
whereas for one period of elliptic motion we have
\begingroup
\allowdisplaybreaks
\small
\begin{align}
\Delta J_\text{ell}(j,\cE)&=\frac{8\,G\,M^2\,\nu^2}{5}\,\Bigg[\frac{15 \pi }{j^4}+\frac{14 \mathcal{E} \pi }{j^2}+\textcolor{1PN}{ \left(\frac{\mathcal{E}^2 \pi  (4283-3976 \nu )}{84 j^2}+\frac{\mathcal{E} \pi  \left(\frac{535}{4}-187 \nu
   \right)}{j^4}-\frac{5 \pi  (-1077+940 \nu )}{48 j^6}\right)}\notag\\
   &\textcolor{2PN}{+\frac{\mathcal{E}^3 \pi  \left(5308-32877 \nu +28308 \nu ^2\right)}{252 j^2}+\frac{5 \mathcal{E}
   \pi  \left(-2999-152946 \nu +106848 \nu ^2\right)}{432 j^6}}\\
   &\textcolor{2PN}{+\frac{\pi  \left(1307683-2782332 \nu +1005480 \nu ^2\right)}{2592 j^8}+\frac{\mathcal{E}^2 \pi 
   \left(-404980-1181889 \nu +1438668 \nu ^2\right)}{1512 j^4}}\notag\\
   &+\textcolor{3PN}{ -\frac{3531 \sqrt{-\mathcal{E}} \pi }{\sqrt{2} j^9}-\frac{10700 \sqrt{2} \sqrt{-\mathcal{E}}
   \mathcal{E} \pi }{3 j^7}-\frac{421366 \sqrt{2} \sqrt{-\mathcal{E}} \mathcal{E}^2 \pi }{315 j^5}}\notag\\
   &\textcolor{3PN}{+\frac{\pi  \left(5567205457+20 \left(-301863524+4719141 \pi ^2\right) \nu
   +2200128840 \nu ^2-371498400 \nu ^3\right)}{322560 j^{10}}}\notag\\
   &\textcolor{3PN}{+\frac{\mathcal{E} \pi}{51840 j^8}  \big(1737906083+160 \left(-7835291+116235 \pi ^2\right) \nu +889119000 \nu
   ^2-277754400 \nu ^3\big)}\notag\\
   &\textcolor{3PN}{-\frac{\mathcal{E}^4 \pi  \left(227005+5072672 \nu -11771892 \nu ^2+9953328 \nu ^3\right)}{44352 j^2}}\notag\\
   &\textcolor{3PN}{-\frac{\mathcal{E}^2 \pi 
   \left(-2082786797+350 \left(126782+23247 \pi ^2\right) \nu -1398279600 \nu ^2+910576800 \nu ^3\right)}{120960 j^6}}\notag\\
   &\textcolor{3PN}{+\frac{\mathcal{E}^3}{j^4} \left(-\frac{1}{16} \left(615
   \pi ^3 \nu \right)+\pi  \left(\frac{90979951}{151200}+\frac{182189 \nu }{108}+\frac{240581 \nu ^2}{84}-3260 \nu ^3\right)\right)}\notag\\
   &\textcolor{3PN}{+\left(\frac{3531 \pi }{2
   j^{10}}+\frac{14231 \mathcal{E} \pi }{3 j^8}+\frac{21614 \mathcal{E}^2 \pi }{7 j^6}+\frac{9844 \mathcal{E}^3 \pi }{35 j^4}\right) \log \left(\frac{(1+\sqrt{1-e_N^2})\mathcal{E}}{2\, (e_N^2-1)}\right)}\Bigg]\,.\notag
\end{align}
\endgroup
The reader will notice the odd vs even factors of $1/j$ in the expansions of the energy and angular momentum. As a result, both entail the same type of analytic continuation, such that we find the latter is as well nicely connected by \eqref{b2bang} to \textcolor{3PN}{3PN} order.

\subsection{Fluxes}
The above values for the total radiated energy and angular momentum were computed from the knowledge of the known PN fluxes in the multipole expansion. However, it is also possible to reverse engineer from the total values, in particular for those obtained from the (on-shell) scattering process. As an example of this inverse problem, let us consider the 3PM  total radiated energy recently computed through various methodologies in \cite{parra2,parra3,max1,max2,janmogul2,4pmeft,4pmzvi,4pmtail},  
\beq
\frac{\Delta E^{(0)}_{\rm hyp}}{M} = -\frac{2\pi \nu^2}{3} \frac{(\gamma^2-1)^2}{\Gamma^4} \chi_{2\epsilon}(\gamma)\,,
\eeq
where the expression for $\chi_{2\epsilon}$ can be found in \cite{4pmeft,4pmtail}. Following the steps outlined in \S\ref{inverse},~ using
 \beq
\int_{J/p_{\infty}}^{\infty} \frac{J^3}{r^4\sqrt{p_{\infty}^2-J^2/r^2}}dr = \frac{\pi}{4} p_{\infty}^2\,,
 \eeq
and
 \beq
{\partial H(r,\bp^2) \over \partial \bp^2 }= {\partial  \over \partial \bp^2} \left(\sqrt{\bp^2+m_1^2}+ \sqrt{\bp^2+m_2^2} +\cdots\right)  = \frac{1}{2\xi E}+{\cal O}(G)\,,
 \eeq
we find 
\beq
 {\cal F}^{(0)}_{ E}(\cE)  =  \frac{2}{\pi} \frac{ \nu \Gamma }{ \xi \, (\gamma^2-1)}  \frac{\Delta E_{\rm hyp}^{(0)} (\cE)}{M} \,,\label{calF0}
\eeq
for the leading coefficient of the PM expansion of the energy flux, see \eqref{f12}.\vskip 4pt 
 
 As we discussed earlier, the energy flux can also be obtained through the pole in the tail Hamiltonian, see \eqref{hpole}. The computation has been carried over in \cite{4pmeft,4pmtail} at 4PM order. As we mentioned, there is a caveat regarding the flux as a function of $(r,\bp^2)$ as in the tail Hamiltonian, and $(r,\cE)$ as in the representation in \S\ref{inverse}. Yet, the mismatch in the PM coefficients of the $G/r$ expansion only matters at higher orders. Hence, we find 
 \bea
 {\cal F}^{(0)}_{ E}(\cE) &=& -\frac{4}{3}\frac{(\gamma^2-1) \nu^3}{ 
\Gamma^3\xi}\chi_{2\epsilon}(\gamma)\,,\label{calF02} 
\eea
 for the 3PM flux, which is equivalent to the result in \eqref{calF0}. Unfortunately, the 3PM value is not sufficient to recover even the leading PN result from the quadrupole formula. However, it is straightforward to compute the required term in the isotropic gauge, yielding\footnote{Notice this turns into a factor of $22\nu^2/15$ when the flux is evaluated as a function of the momentum \cite{4pmeft}.}
 \beq
  {\cal F}^{(1)}_{ E}(\cE) = \frac{34\nu^2}{3} + {\cal O}({\cal E})\,.
  \eeq

\subsection{Aligned-spin configurations}
Using the results for the spin-dependent dynamical effects to next-to-leading order \cite{nrgrs,nrgrso,nrgrss,nrgrs2,srad,amps}, recently combined in \cite{Cho:2021mqw} to compute the associated fluxes (see also \cite{Cho:2022syn}), we derived the radiated energy and angular momentum for aligned-spin configurations. 
To the extent of our knowledge, the following results are reported here for the first time.\vskip 4pt For one period of elliptic-like motion we find to next-to-leading PN order and quadratic order in the spins,
\begingroup
\allowdisplaybreaks
\small
\begin{align}
  &\frac{\Delta E_\textrm{ell}(\ell,\tilde{a}_\pm,\cE)}{M \pi \nu^2}=\\
  &\quad\frac{1}{\ell^4}\left\{
    -\frac{4}{5} \cE^3 [33 \tilde{a}_- \Delta +97 \tilde{a}_+]
    +\frac{1}{70} \cE^4 [\nu  (7504 \tilde{a}_- \Delta +20048 \tilde{a}_+)-4869 \tilde{a}_- \Delta -15806 \tilde{a}_+]
  \right\}\nn\\
  &\quad+\frac{1}{\ell^5}
  \begin{multlined}[t]
    \bigg\{
      \frac{2}{5} \cE^3 \left[\tilde{a}_-^2 (59 \kappa_+-109)+118 \tilde{a}_- \tilde{a}_+ \kappa_-+59 \tilde{a}_+^2 (\kappa_++2)\right]\\
      +\frac{1}{280} \cE^4 \left[\nu  \left(-56 \tilde{a}_-^2 (531 \kappa_+-517)-59472 \tilde{a}_- \tilde{a}_+ \kappa_--29736 \tilde{a}_+^2 (\kappa_++2)\right)\right.\\
      +\tilde{a}_-^2 (609 \Delta  \kappa_-+26643 \kappa_+-38138)+6 \tilde{a}_- \tilde{a}_+ (203 \Delta  \kappa_++6104 \Delta +8881 \kappa_-)\\
      \left.+\tilde{a}_+^2 (609 \Delta  \kappa_-+26643 \kappa_++49478)\right]
    \bigg\}
  \end{multlined}\nn\\
  &\quad+\frac{1}{\ell^6}
  \begin{multlined}[t]
    \bigg\{
      -12 \cE^2 [34 \tilde{a}_- \Delta +145 \tilde{a}_+]
      +\frac{1}{42} \cE^3 [\nu  (127456 \tilde{a}_- \Delta +509152 \tilde{a}_+)-114753 \tilde{a}_- \Delta -372729 \tilde{a}_+]
    \bigg\}
  \end{multlined}\nn\\
  &\quad+\frac{1}{\ell^7}
  \begin{multlined}[t]
    \bigg\{
      \cE^2 \left[334 \tilde{a}_-^2 \kappa_+-635 \tilde{a}_-^2+668 \tilde{a}_- \tilde{a}_+ \kappa_-+334 \tilde{a}_+^2 \kappa_++668 \tilde{a}_+^2\right]\\
      +\frac{1}{84} \cE^3 \left[\nu  \left(-112 \tilde{a}_-^2 (2037 \kappa_+-1214)-456288 \tilde{a}_- \tilde{a}_+ \kappa_--228144 \tilde{a}_+^2 (\kappa_++2)\right)\right.\\
      +\tilde{a}_-^2 (22827 \Delta  \kappa_-+215028 \kappa_+-328472)+6 \tilde{a}_- \tilde{a}_+ (7609 \Delta  \kappa_++93156 \Delta +71676 \kappa_-)\\
      \left.+\tilde{a}_+^2 (22827 \Delta  \kappa_-+215028 \kappa_++1010300)\right]
    \bigg\}
  \end{multlined}\nn\\
  &\quad+\frac{1}{\ell^8}
  \begin{multlined}[t]
    \bigg\{
      -833 \cE [\tilde{a}_- \Delta +5 \tilde{a}_+]
      +\frac{1}{12} \cE^2 [\nu  (136500 \tilde{a}_- \Delta +665364 \tilde{a}_+)-187318 \tilde{a}_- \Delta -651175 \tilde{a}_+]
    \bigg\}
  \end{multlined}\nn\\
  &\quad+\frac{1}{\ell^9}
  \begin{multlined}[t]
    \bigg\{
      \frac{7}{6} \cE \left[\tilde{a}_-^2 (539 \kappa_+-1039)+1078 \tilde{a}_- \tilde{a}_+ \kappa_-+539 \tilde{a}_+^2 (\kappa_++2)\right]\\
      +\frac{7}{48} \cE^2 \left[\nu  \left(-4 \tilde{a}_-^2 (16099 \kappa_+-1099)-128792 \tilde{a}_- \tilde{a}_+ \kappa_--64396 \tilde{a}_+^2 (\kappa_++2)\right)\right.\\
      +\tilde{a}_-^2 (9693 \Delta  \kappa_-+83685 \kappa_+-134138)+2 \tilde{a}_- \tilde{a}_+ (9693 \Delta  \kappa_++130688 \Delta +83685 \kappa_-)\\
      \left.+3 \tilde{a}_+^2 (3231 \Delta  \kappa_-+27895 \kappa_++201638)\right]
    \bigg\}
  \end{multlined}\nn\\
  &\quad+\frac{1}{\ell^{10}}
  \begin{multlined}[t]
    \bigg\{
      -\frac{21}{5} [93 \tilde{a}_- \Delta +506 \tilde{a}_+]
      +\frac{3}{40} \cE [\nu  (162988 \tilde{a}_- \Delta +895188 \tilde{a}_+)-328441 \tilde{a}_- \Delta -1285277 \tilde{a}_+]
    \bigg\}
  \end{multlined}\nn\\
  &\quad+\frac{1}{\ell^{11}}
  \begin{multlined}[t]
    \bigg\{
      \frac{63}{20} \left[\tilde{a}_-^2 (88 \kappa_+-171)+176 \tilde{a}_- \tilde{a}_+ \kappa_-+88 \tilde{a}_+^2 (\kappa_++2)\right]\\
      +\frac{3}{80} \cE \left[\nu  \left(-196 \tilde{a}_-^2 (1296 \kappa_++761)-508032 \tilde{a}_- \tilde{a}_+ \kappa_--254016 \tilde{a}_+^2 (\kappa_++2)\right)\right.\\
      +\tilde{a}_-^2 (48321 \Delta  \kappa_-+470006 \kappa_+-775904)+2 \tilde{a}_- \tilde{a}_+ (48321 \Delta  \kappa_++770532 \Delta +470006 \kappa_-)\\
      \left.+\tilde{a}_+^2 (48321 \Delta  \kappa_-+470006 \kappa_++4121960)\right]
    \bigg\}
  \end{multlined}\nn\\
  &\quad+\frac{1}{\ell^{12}}
  \begin{multlined}[t]
    \bigg\{
      \frac{11 (\nu  (387296 \tilde{a}_- \Delta +2308880 \tilde{a}_+)-1097925 \tilde{a}_- \Delta -4756668 \tilde{a}_+)}{1120}
    \bigg\}
  \end{multlined}\nn\\
  &\quad+\frac{1}{\ell^{13}}
  \begin{multlined}[t]
    \bigg\{
      \frac{11}{4480}\left[\nu  \left(-112 \tilde{a}_-^2 (10347 \kappa_++14269)-2317728 \tilde{a}_- \tilde{a}_+ \kappa_--1158864 \tilde{a}_+^2 (\kappa_++2)\right)\right.\\
        +\tilde{a}_-^2 (262227 \Delta  \kappa_-+2978061 \kappa_+-5005550)+6 \tilde{a}_- \tilde{a}_+ (87409 \Delta  \kappa_++1644440 \Delta +992687 \kappa_-)\\
        \left.+\tilde{a}_+^2 (262227 \Delta  \kappa_-+2978061 \kappa_++28885322)\right]
    \bigg\}
  \end{multlined}\nn
\end{align}
\endgroup
The computation for unbound orbits is significantly more involved. However, after some massaging it can be written as follows:
\beq
\Delta E_\text{hyp}(\ell,\tilde{a}_\pm,\cE)= +\frac{\Delta E_\text{ell}(\ell,\tilde{a}_\pm,\cE)}{\pi}\,\cos^{-1}\left(\frac{-1}{e_N}\right) + \Delta E^{(\rm even)}_\text{hyp}(\ell,\tilde{a}_\pm,\cE)\,,
\eeq
where $\Delta E^{(\rm even)}_\text{hyp}$ is even under $J \to -J$ (and $e \to -e$). Hence, noticing that $\Delta E_\text{ell}$ is odd under $J$-parity,  and using \eqref{arcos}, the B2B map in \eqref{b2ben} is obeyed.\vskip 4pt The explicit expression for the (lengthier) even term is given by:

\begingroup
\allowdisplaybreaks
\small
\begin{align}
  &\frac{\sqrt{2}e_N^4\Delta E_\textrm{hyp}^\textrm{(even)}(\ell,\tilde{a}_\pm,\cE)}{M\nu^2}=\\
  &\quad\frac{1}{\ell}
  \begin{multlined}[t]
    \bigg\{
      -\frac{16}{75} \cE^{9/2} [5713 \tilde{a}_- \Delta +21521 \tilde{a}_+]\\
      +\frac{2 \cE^{11/2} [\nu  (38147088 \tilde{a}_- \Delta +131474448 \tilde{a}_+)-28359001 \tilde{a}_- \Delta -93954742 \tilde{a}_+]}{11025}
    \bigg\}
  \end{multlined}\nn\\
  &\quad\frac{1}{\ell^2}
  \begin{multlined}[t]
    \bigg\{
      \frac{8}{225} \cE^{9/2} \left[\tilde{a}_-^2 (29281 \kappa_+-55127)+58562 \tilde{a}_- \tilde{a}_+ \kappa_-+29281 \tilde{a}_+^2 (\kappa_++2)\right]\\
     +\frac{\cE^{11/2}}{22050} \left[\nu  \left(-56 \tilde{a}_-^2 (2554367 \kappa_+-2034713)-286089104 \tilde{a}_- \tilde{a}_++\kappa_--143044552 \tilde{a}_+^2 (\kappa_++2)\right)\right.\\
       +\tilde{a}_-^2 (9872037 \Delta  \kappa_-+123495327 \kappa_+-183275618)\\
       +2 \tilde{a}_- \tilde{a}_+ (9872037 \Delta  \kappa_++130785704 \Delta +123495327 \kappa_-)\\
       \left.+9 \tilde{a}_+^2 (1096893 \Delta  \kappa_-+13721703 \kappa_++44532590)\right]
    \bigg\}
  \end{multlined}\nn\\
  &\quad\frac{1}{\ell^3}
  \begin{multlined}[t]
    \bigg\{
      -\frac{8}{75} \cE^{7/2} [54851 \tilde{a}_- \Delta +252517 \tilde{a}_+]\\
      +\frac{\cE^{9/2} [\nu  (630281904 \tilde{a}_- \Delta +2806389936 \tilde{a}_+)-702186917 \tilde{a}_- \Delta -2332888208 \tilde{a}_+]}{11025}
    \bigg\}
  \end{multlined}\nn\\
  &\quad\frac{1}{\ell^4}
  \begin{multlined}[t]
    \bigg\{
      \frac{4}{225} \cE^{7/2} \left[\tilde{a}_-^2 (259817 \kappa_+-497014)+519634 \tilde{a}_- \tilde{a}_+ \kappa_-+259817 \tilde{a}_+^2 (\kappa_++2)\right]\\
      +\frac{\cE^{9/2}}{44100} \left[\nu  \left(-56 \tilde{a}_-^2 (38656801 \kappa_+-13673167)-4329561712 \tilde{a}_- \tilde{a}_+ \kappa_--2164780856 \tilde{a}_+^2 (\kappa_++2)\right)\right.\\
        +\tilde{a}_-^2 (271723095 \Delta  \kappa_-+2371844217 \kappa_+-3718678126)\\
        +2 \tilde{a}_- \tilde{a}_+ (271723095 \Delta  \kappa_++3450724648 \Delta +2371844217 \kappa_-)\\
        \left.+3 \tilde{a}_+^2 (90574365 \Delta  \kappa_-+790614739 \kappa_++4794170902)\right]
    \bigg\}
  \end{multlined}\nn\\
  &\quad\frac{1}{\ell^5}
  \begin{multlined}[t]
    \bigg\{
    -\frac{4}{75} \cE^{5/2} [150673 \tilde{a}_- \Delta +755891 \tilde{a}_+]\\
    +\frac{\cE^{7/2} [\nu  (1426875492 \tilde{a}_- \Delta +7196110404 \tilde{a}_+)-2221381823 \tilde{a}_- \Delta -8090706680 \tilde{a}_+]}{11025}
    \bigg\}
  \end{multlined}\nn\\
  &\quad\frac{1}{\ell^6}
  \begin{multlined}[t]
    \bigg\{
      \frac{2}{225} \cE^{5/2} \left[\tilde{a}_-^2 (679831 \kappa_+-1310552)+1359662 \tilde{a}_- \tilde{a}_+ \kappa_-+679831 \tilde{a}_+^2 (\kappa_++2)\right]\\
      +\frac{\cE^{7/2}}{44100} \left[\nu  \left(-28 \tilde{a}_-^2 (165710591 \kappa_++18223105)-9279793096 \tilde{a}_- \tilde{a}_+ \kappa_--4639896548 \tilde{a}_+^2 (\kappa_++2)\right)\right.\\
        +\tilde{a}_-^2 (745894737 \Delta  \kappa_-+6737032059 \kappa_+-10910108374)\\
        +2 \tilde{a}_- \tilde{a}_+ (745894737 \Delta  \kappa_++10682364952 \Delta +6737032059 \kappa_-)\\
        \left.+3 \tilde{a}_+^2 (248631579 \Delta  \kappa_-+2245677353 \kappa_++17422730002)\right]
    \bigg\}
  \end{multlined}\nn\\
  &\quad\frac{1}{\ell^7}
  \begin{multlined}[t]
    \bigg\{
      -14 \cE^{3/2} [305 \tilde{a}_- \Delta +1607 \tilde{a}_+]\\
      +\frac{\cE^{5/2} [\nu  (126497224 \tilde{a}_- \Delta +688003120 \tilde{a}_+)-253520875 \tilde{a}_- \Delta -998019442 \tilde{a}_+]}{1050}
    \bigg\}
  \end{multlined}\nn\\
  &\quad\frac{1}{\ell^8}
  \begin{multlined}[t]
    \bigg\{
      \frac{7}{3} \cE^{3/2} \left[\tilde{a}_-^2 (1331 \kappa_+-2578)+2662 \tilde{a}_- \tilde{a}_+ \kappa_-+1331 \tilde{a}_+^2 (\kappa_++2)\right]\\
      +\frac{\cE^{5/2}}{4200} \left[\nu  \left(-28 \tilde{a}_-^2 (14163997 \kappa_++7912369)-793183832 \tilde{a}_- \tilde{a}_+ \kappa_--396591916 \tilde{a}_+^2 (\kappa_++2)\right)\right.\\
        +3 \tilde{a}_-^2 (24716321 \Delta  \kappa_-+242351703 \kappa_+-399754150)\\
        +2 \tilde{a}_- \tilde{a}_+ (74148963 \Delta  \kappa_++1186217480 \Delta +727055109 \kappa_-)\\
        \left.+\tilde{a}_+^2 (74148963 \Delta  \kappa_-+727055109 \kappa_++6340031818)\right]
    \bigg\}
  \end{multlined}\nn\\
  &\quad\frac{1}{\ell^9}
  \begin{multlined}[t]
    \bigg\{
      -\frac{42}{5} \sqrt{\cE} [93 \tilde{a}_- \Delta +506 \tilde{a}_+]\\
      +\frac{1}{840} \cE^{3/2} [\nu  (41837768 \tilde{a}_- \Delta +239782088 \tilde{a}_+)-101769441 \tilde{a}_- \Delta -423561642 \tilde{a}_+]
    \bigg\}
  \end{multlined}\nn\\
  &\quad\frac{1}{\ell^{10}}
  \begin{multlined}[t]
    \bigg\{
      \frac{63}{10} \sqrt{\cE} \left[\tilde{a}_-^2 (88 \kappa_+-171)+176 \tilde{a}_- \tilde{a}_+ \kappa_-+88 \tilde{a}_+^2 (\kappa_++2)\right]\\
      +\frac{\cE^{3/2}}{3360} \left[\nu  \left(-112 \tilde{a}_-^2 (1140621 \kappa_++1120396)-255499104 \tilde{a}_- \tilde{a}_+ \kappa_--127749552 \tilde{a}_+^2 (\kappa_++2)\right)\right.\\
        +\tilde{a}_-^2 (26599377 \Delta  \kappa_-+282234867 \kappa_+-470833058)\\
        +6 \tilde{a}_- \tilde{a}_+ (8866459 \Delta  \kappa_++155168888 \Delta +94078289 \kappa_-)\\
        \left.+\tilde{a}_+^2 (26599377 \Delta  \kappa_-+282234867 \kappa_++2627426630)\right]
    \bigg\}
  \end{multlined}\nn\\
  &\quad\frac{1}{\ell^{11}}
  \begin{multlined}[t]
    \bigg\{
      \frac{11}{560} \sqrt{\cE} [\nu  (387296 \tilde{a}_- \Delta +2308880 \tilde{a}_+)-1097925 \tilde{a}_- \Delta -4756668 \tilde{a}_+]
    \bigg\}
  \end{multlined}\nn\\
  &\quad\frac{1}{\ell^{12}}
  \begin{multlined}[t]
    \bigg\{
      \frac{11 \sqrt{\cE}}{2240} \left[\nu  \left(-112 \tilde{a}_-^2 (10347 \kappa_++14269)-2317728 \tilde{a}_- \tilde{a}_+ \kappa_--1158864 \tilde{a}_+^2 (\kappa_++2)\right)\right.\\
        +\tilde{a}_-^2 (262227 \Delta  \kappa_-+2978061 \kappa_+-5005550)\\
        +6 \tilde{a}_- \tilde{a}_+ (87409 \Delta  \kappa_++1644440 \Delta +992687 \kappa_-)\\
        \left.+\tilde{a}_+^2 (262227 \Delta  \kappa_-+2978061 \kappa_++28885322)\right]
    \bigg\}
  \end{multlined}\nn
\end{align}
\endgroup
Something similar occurs for the angular momentum, where we find

\begingroup
\allowdisplaybreaks
\small
\begin{align}
  &\frac{\Delta J_\textrm{ell}(\ell,\tilde{a}_\pm,\cE)}{\pi G M^2\nu^2}=\\
  &\quad\frac{1}{\ell^3}\left\{
    -\frac{8}{15} \cE^2 [52 \tilde{a}_- \Delta +245 \tilde{a}_+]
    +\frac{2}{15} \cE^3 [\nu  (980 \tilde{a}_- \Delta +3680 \tilde{a}_+)-898 \tilde{a}_- \Delta -2159 \tilde{a}_+]
  \right\}\nn\\
  &\quad+\frac{1}{\ell^4}
  \begin{multlined}[t]
    \bigg\{
      \frac{6}{5} \cE^2 \left[\tilde{a}_-^2 (21 \kappa_+-41)+42 \tilde{a}_- \tilde{a}_+ \kappa_-+21 \tilde{a}_+^2 (\kappa_++2)\right]\\
      +\frac{1}{70} \cE^3 \left[\nu  \left(-28 \tilde{a}_-^2 (311 \kappa_+-369)-17416 \tilde{a}_- \tilde{a}_+ \kappa_--8708 \tilde{a}_+^2 (\kappa_++2)\right)\right.\\
        +\tilde{a}_-^2 (371 \Delta  \kappa_-+9419 \kappa_+-16514)+2 \tilde{a}_- \tilde{a}_+ (371 \Delta  \kappa_++7868 \Delta +9419 \kappa_-)\\
        \left.+\tilde{a}_+^2 (371 \Delta  \kappa_-+9419 \kappa_++18530)\right]
    \bigg\}
  \end{multlined}\nn\\
  &\quad+\frac{1}{\ell^5}
  \begin{multlined}[t]
    \bigg\{
      -\frac{8}{5} \cE [95 \tilde{a}_- \Delta +468 \tilde{a}_+]
      +\frac{1}{35} \cE^2 [\nu  (55832 \tilde{a}_- \Delta +242872 \tilde{a}_+)-58346 \tilde{a}_- \Delta -166243 \tilde{a}_+]
    \bigg\}
  \end{multlined}\nn\\
  &\quad+\frac{1}{\ell^6}
  \begin{multlined}[t]
    \bigg\{
      6 \cE \left[\tilde{a}_-^2 (20 \kappa_+-39)+40 \tilde{a}_- \tilde{a}_+ \kappa_-+20 \tilde{a}_+^2 (\kappa_++2)\right]\\
      +\frac{1}{28} \cE^2 \left[\nu  \left(-28 \tilde{a}_-^2 (1324 \kappa_+-975)-74144 \tilde{a}_- \tilde{a}_+ \kappa_--37072 \tilde{a}_+^2 (\kappa_++2)\right)\right.\\
        +\tilde{a}_-^2 (4319 \Delta  \kappa_-+38618 \kappa_+-64972)+2 \tilde{a}_- \tilde{a}_+ (4319 \Delta  \kappa_++49308 \Delta +38618 \kappa_-)\\
        \left.+\tilde{a}_+^2 (4319 \Delta  \kappa_-+38618 \kappa_++190300)\right]
    \bigg\}
  \end{multlined}\nn\\
  &\quad+\frac{1}{\ell^7}
  \begin{multlined}[t]
    \bigg\{
      -\frac{2}{3} [168 \tilde{a}_- \Delta +901 \tilde{a}_+]
      +\frac{1}{6} \cE [\nu  (17536 \tilde{a}_- \Delta +88648 \tilde{a}_+)-26100 \tilde{a}_- \Delta -87115 \tilde{a}_+]
    \bigg\}
  \end{multlined}\nn\\
  &\quad+\frac{1}{\ell^8}
  \begin{multlined}[t]
    \bigg\{
      \frac{7}{2} \left[\tilde{a}_-^2 (23 \kappa_+-45)+46 \tilde{a}_- \tilde{a}_+ \kappa_-+23 \tilde{a}_+^2 (\kappa_++2)\right]\\
      +\frac{1}{24} \cE \left[\nu  \left(-56 \tilde{a}_-^2 (961 \kappa_+-154)-107632 \tilde{a}_- \tilde{a}_+ \kappa_--53816 \tilde{a}_+^2 (\kappa_++2)\right)\right.\\
        +\tilde{a}_-^2 (8869 \Delta  \kappa_-+74819 \kappa_+-126006)+2 \tilde{a}_- \tilde{a}_+ (8869 \Delta  \kappa_++113260 \Delta +74819 \kappa_-)\\
        \left.+\tilde{a}_+^2 (8869 \Delta  \kappa_-+74819 \kappa_++549338)\right]
    \bigg\}
  \end{multlined}\nn\\
  &\quad+\frac{1}{\ell^9}
  \begin{multlined}[t]
    \bigg\{
      \frac{1}{60} [\nu  (74676 \tilde{a}_- \Delta +418432 \tilde{a}_+)-158892 \tilde{a}_- \Delta -615951 \tilde{a}_+]
    \bigg\}
  \end{multlined}\nn\\
  &\quad+\frac{1}{\ell^{10}}
  \begin{multlined}[t]
    \bigg\{
      \frac{3}{80} \left[\nu  \left(-112 \tilde{a}_-^2 (216 \kappa_++115)-48384 \tilde{a}_- \tilde{a}_+ \kappa_--24192 \tilde{a}_+^2 (\kappa_++2)\right)\right.\\
        +\tilde{a}_-^2 (4921 \Delta  \kappa_-+47168 \kappa_+-80140)+2 \tilde{a}_- \tilde{a}_+ (4921 \Delta  \kappa_++75348 \Delta +47168 \kappa_-)\\
        \left.+\tilde{a}_+^2 (4921 \Delta  \kappa_-+47168 \kappa_++416112)\right]
    \bigg\}
  \end{multlined}\nn
\end{align}
\endgroup
On the other hand, after similar manipulations, we arrive at
\beq
\Delta J_\text{hyp}(\ell,\tilde{a}_\pm,\cE)= +\frac{\Delta J_\text{ell}(\ell,\tilde{a}_\pm,\cE)}{\pi}\,\cos^{-1}\left(\frac{-1}{e_N}\right) + \Delta J^{(\rm odd)}_\text{hyp}(\ell,\tilde{a}_\pm,\cE)\,,
\eeq
where $\Delta J^{(\rm even)}_\text{hyp}$ is odd under $J \to -J$ (or $e \to -e$). Once again, since in this case $\Delta J_\text{ell}$ is even under $J$-parity, the B2B correspondence in \eqref{b2bang}~applies.\vskip 4pt The expression for the odd term is given by:
\begingroup
\allowdisplaybreaks
\small
\begin{align}
  &\frac{\sqrt{2}e_N^4\Delta J_\textrm{hyp}^\textrm{(odd)}(\ell,\tilde{a}_\pm,\cE)}{G M^2\nu^2}=\\
  &\quad\ell^2\left\{
    -\frac{128}{5} \cE^{9/2} [\tilde{a}_- \Delta +7 \tilde{a}_+]
    +\frac{32}{35} \cE^{11/2} [\nu  (63 \tilde{a}_- \Delta +357 \tilde{a}_+)-109 \tilde{a}_- \Delta -203 \tilde{a}_+]
  \right\}\nn\\
  &\quad+\ell
  \begin{multlined}[t]
  \bigg\{
    \frac{128}{5} \cE^{9/2} \left[\tilde{a}_-^2 (\kappa_+-2)+2 \tilde{a}_- \tilde{a}_+ \kappa_-+\tilde{a}_+^2 (\kappa_++2)\right]\\
    -\frac{32}{35} \cE^{11/2} \left[\nu  \left(7 \tilde{a}_-^2 (11 \kappa_+-14)+154 \tilde{a}_- \tilde{a}_+ \kappa_-+77 \tilde{a}_+^2 (\kappa_++2)\right)\right.\\
      \left.+\tilde{a}_-^2 (232-123 \kappa_+)-2 \tilde{a}_- \tilde{a}_+ (98 \Delta +123 \kappa_-)-3 \tilde{a}_+^2 (41 \kappa_++12)\right]
  \bigg\}
  \end{multlined}\nn\\
  &\quad+\ell^0
  \begin{multlined}[t]
  \bigg\{
    -\frac{32}{45} \cE^{7/2} [978 \tilde{a}_- \Delta +4675 \tilde{a}_+]\\
    +\frac{8 \cE^{9/2} [\nu  (1009106 \tilde{a}_- \Delta +4045202 \tilde{a}_+)-935472 \tilde{a}_- \Delta -2512391 \tilde{a}_+]}{1575}
  \bigg\}
  \end{multlined}\nn\\
  &\quad+\frac{1}{\ell}
  \begin{multlined}[t]
  \bigg\{
    \frac{8}{15} \cE^{7/2} \left[\tilde{a}_-^2 (1091 \kappa_+-2127)+2182 \tilde{a}_- \tilde{a}_+ \kappa_-+1091 \tilde{a}_+^2 (\kappa_++2)\right]\\
    +\frac{2 \cE^{9/2}}{1575} \left[\nu  \left(-28 \tilde{a}_-^2 (126061 \kappa_+-121315)-7059416 \tilde{a}_- \tilde{a}_+ \kappa_--3529708 \tilde{a}_+^2 (\kappa_++2)\right)\right.\\
      +\tilde{a}_-^2 (309449 \Delta  \kappa_-+3479857 \kappa_+-5918310)
      +2 \tilde{a}_- \tilde{a}_+ (309449 \Delta  \kappa_++3785012 \Delta +3479857 \kappa_-)\\
      \left.+\tilde{a}_+^2 (309449 \Delta  \kappa_-+3479857 \kappa_++12238918)\right]
  \bigg\}
  \end{multlined}\nn\\
  &\quad+\frac{1}{\ell^2}
  \begin{multlined}[t]
  \bigg\{
    -\frac{16}{45} \cE^{5/2} [4350 \tilde{a}_- \Delta +21983 \tilde{a}_+]\\
\frac{16 \cE^{7/2} [\nu  (2150743 \tilde{a}_- \Delta +10018946 \tilde{a}_+)-2626776 \tilde{a}_- \Delta -8039878 \tilde{a}_+]}{1575}
  \bigg\}
  \end{multlined}\nn\\
  &\quad+\frac{1}{\ell^3}
  \begin{multlined}[t]
  \bigg\{
    \frac{4}{15} \cE^{5/2} \left[\tilde{a}_-^2 (4477 \kappa_+-8739)+8954 \tilde{a}_- \tilde{a}_+ \kappa_-+4477 \tilde{a}_+^2 (\kappa_++2)\right]\\
    +\frac{\cE^{7/2}}{1575} \left[\nu  \left(-28 \tilde{a}_-^2 (986447 \kappa_+-501590)-55241032 \tilde{a}_- \tilde{a}_+ \kappa_--27620516 \tilde{a}_+^2 (\kappa_++2)\right)\right.\\
      +\tilde{a}_-^2 (3765328 \Delta  \kappa_-+32458829 \kappa_+-54619230)\\
      +2 \tilde{a}_- \tilde{a}_+ (3765328 \Delta  \kappa_++45132304 \Delta +32458829 \kappa_-)\\
      \left.+\tilde{a}_+^2 (3765328 \Delta  \kappa_-+32458829 \kappa_++196050926)\right]
  \bigg\}
  \end{multlined}\nn\\
  &\quad+\frac{1}{\ell^4}
  \begin{multlined}[t]
  \bigg\{
    -\frac{8}{45} \cE^{3/2} [5910 \tilde{a}_- \Delta +30949 \tilde{a}_+]\\
    +\frac{4 \cE^{5/2} [\nu  (11019148 \tilde{a}_- \Delta +55964216 \tilde{a}_+)-17180511 \tilde{a}_- \Delta -59099908 \tilde{a}_+]}{1575}
  \bigg\}
  \end{multlined}\nn\\
  &\quad+\frac{1}{\ell^5}
  \begin{multlined}[t]
  \bigg\{
    \frac{2}{3} \cE^{3/2} \left[\tilde{a}_-^2 (1165 \kappa_+-2277)+2330 \tilde{a}_- \tilde{a}_+ \kappa_-+1165 \tilde{a}_+^2 (\kappa_++2)\right]\\
    +\frac{\cE^{5/2}}{3150} \left[\nu  \left(-28 \tilde{a}_-^2 (2415106 \kappa_+-257035)-135245936 \tilde{a}_- \tilde{a}_+ \kappa_--67622968 \tilde{a}_+^2 (\kappa_++2)\right)\right.\\
      +\tilde{a}_-^2 (11212334 \Delta  \kappa_-+97928347 \kappa_+-165264510)\\
      +2 \tilde{a}_- \tilde{a}_+ (11212334 \Delta  \kappa_++148172192 \Delta +97928347 \kappa_-)\\
      \left.+\tilde{a}_+^2 (11212334 \Delta  \kappa_-+97928347 \kappa_++733208698)\right]
  \bigg\}
  \end{multlined}\nn\\
  &\quad+\frac{1}{\ell^6}
  \begin{multlined}[t]
  \bigg\{
    -\frac{4}{3} \sqrt{\cE} [168 \tilde{a}_- \Delta +901 \tilde{a}_+]\\
    +\frac{4}{9} \cE^{3/2} [\nu  (31821 \tilde{a}_- \Delta +171094 \tilde{a}_+)-59298 \tilde{a}_- \Delta -219324 \tilde{a}_+]
  \bigg\}
  \end{multlined}\nn\\
  &\quad+\frac{1}{\ell^7}
  \begin{multlined}[t]
  \bigg\{
    7 \sqrt{\cE} \left[\tilde{a}_-^2 (23 \kappa_+-45)+46 \tilde{a}_- \tilde{a}_+ \kappa_-+23 \tilde{a}_+^2 (\kappa_++2)\right]\\
    +\frac{1}{12} \cE^{3/2} \left[\nu  \left(-56 \tilde{a}_-^2 (2257 \kappa_++536)-252784 \tilde{a}_- \tilde{a}_+ \kappa_--126392 \tilde{a}_+^2 (\kappa_++2)\right)\right.\\
      +\tilde{a}_-^2 (23632 \Delta  \kappa_-+216323 \kappa_+-366426)+2 \tilde{a}_- \tilde{a}_+ (23632 \Delta  \kappa_++339304 \Delta +216323 \kappa_-)\\
      \left.+\tilde{a}_+^2 (23632 \Delta  \kappa_-+216323 \kappa_++1797674)\right]
  \bigg\}
  \end{multlined}\nn\\
  &\quad+\frac{1}{\ell^8}
  \begin{multlined}[t]
  \bigg\{
    \frac{1}{30} \sqrt{\cE} [\nu  (74676 \tilde{a}_- \Delta +418432 \tilde{a}_+)-158892 \tilde{a}_- \Delta -615951 \tilde{a}_+]
  \bigg\}
  \end{multlined}\nn\\
  &\quad+\frac{1}{\ell^9}
  \begin{multlined}[t]
  \bigg\{
    \frac{3}{40} \sqrt{\cE} \left[\nu  \left(-112 \tilde{a}_-^2 (216 \kappa_++115)-48384 \tilde{a}_- \tilde{a}_+ \kappa_--24192 \tilde{a}_+^2 (\kappa_++2)\right)\right.\\
      +\tilde{a}_-^2 (4921 \Delta  \kappa_-+47168 \kappa_+-80140)+2 \tilde{a}_- \tilde{a}_+ (4921 \Delta  \kappa_++75348 \Delta +47168 \kappa_-)\\
      \left.+\tilde{a}_+^2 (4921 \Delta  \kappa_-+47168 \kappa_++416112)\right]
  \bigg\}
  \end{multlined}\nn
\end{align}
\endgroup
Notice in both cases some of the terms in the hyperbolic result do not allow for a smooth analytic continuation in the binding energy. However, all of these pieces nicely cancel out in the B2B dictionary of \eqref{b2ben} or \eqref{b2bang}.

\subsection{Local-in-time}
We move on now onto the conservative sector and local-in-time effects. We concentrate on the \textcolor{4PN}{4PN} dynamics which has been established by independent derivations \cite{Damour:2014jta,Jaranowski:2015lha,Bernard:2017bvn,Marchand:2017pir,tail,apparent,nrgr4pn1,nrgr4pn2}.\footnote{The conservative \textcolor{5PN}{5PN} dynamics was recently reported in \cite{5pnfin1,5pnfin2}, see also \cite{hered,foffa3,foffa4,bini0,bini1,Bini:2021gat}, and partial values to \textcolor{6PN}{6PN} order in \cite{binidam1,binidam2}. Despite some disagreements in the literature, notably for memory terms, we checked the B2B correspondence remains valid regardless of the specific values associated with local-in-time dynamics.}\vskip 4pt Introducing the split between local- and non-local-in-time contributions,
\beq
\frac{\chi}{2} = \sum_{j=1}^{\infty} \frac{1}{j^n}\left(\chi^{(n)}_{j,\rm loc} (v_\infty)+\chi^{(n)}_{j,\rm nloc} (v_\infty)\right)\,,
\eeq
where $v_{\infty}\equiv \sqrt{\gamma^2-1}$ serves as an expansion parameter, we have 
 \beq
 \allowdisplaybreaks
  \begin{aligned}
    \chi_{j,\textrm{loc}}^{(1)} &= \frac{1}{v_\infty}+\textcolor{1PN}{2 v_\infty}\,,\\
    \chi_{j,\textrm{loc}}^{(2)} &=\textcolor{1PN}{\frac{3 \pi }{2}}
    +\textcolor{2PN}{\left[\frac{15 \pi }{8}-\frac{3 \pi  \nu }{4}\right] v_\infty^2}
    +\textcolor{3PN}{\left[\frac{9\pi  \nu ^2}{16}-\frac{3 \pi  \nu }{4}\right] v_\infty^4}
    +\textcolor{4PN}{\left[-\frac{15 \pi  \nu ^3}{32}+\frac{27 \pi  \nu ^2}{64}+\frac{9 \pi  \nu }{64}\right] v_\infty^6}\\
    \chi_{j,\textrm{loc}}^{(3)} &= -\frac{1}{3 v_\infty^3}
    +\textcolor{1PN}{\frac{4}{v_\infty}}
    +\textcolor{2PN}{[24-8 \nu ] v_\infty}
    +\textcolor{3PN}{\left[8 \nu ^2-36 \nu +\frac{64}{3}\right] v_\infty^3}+\textcolor{4PN}{\left[-8 \nu ^3+34 \nu ^2-\frac{91 \nu }{5}\right] v_\infty^5}\\
    \chi_{j,\textrm{loc}}^{(4)} &= \textcolor{2PN}{\left[-\frac{15 \pi  \nu }{4}+\frac{105 \pi }{8}\right]}
    +\textcolor{3PN}{\left[\frac{45 \pi  \nu ^2}{8}+\left(\frac{123 \pi ^3}{256}-\frac{109 \pi }{2}\right) \nu +\frac{315 \pi }{8}\right] v_\infty^2}\\
    &\quad+\textcolor{4PN}{\left[-\frac{225 \pi  \nu ^3}{32}-\frac{3}{512} \pi  \left(123 \pi ^2-12872\right) \nu ^2+\frac{\pi  \left(100803 \pi ^2-5016832\right) \nu }{49152}+\frac{3465 \pi }{128}\right] v_\infty^4}\\
       \chi_{j,\textrm{loc}}^{(5)} &=\frac{1}{5 v_\infty^5}
    -\textcolor{1PN}{\frac{2}{v_\infty^3}}
    +\textcolor{2PN}{\frac{32-8 \nu }{v_\infty}}
    +\textcolor{3PN}{\left[24 \nu^2+\left(\frac{41 \pi ^2}{8}-\frac{1168}{3}\right) \nu +320\right] v_\infty}\\
    &\quad+\textcolor{4PN}{\left[-40 \nu ^3+\left(\frac{7342}{9}-\frac{287 \pi ^2}{24}\right) \nu ^2+\left(\frac{5069 \pi ^2}{144}-\frac{227059}{135}\right) \nu +640\right] v_\infty^3}\\
  \end{aligned}
  \eeq
From here, and using  the algebraic relations discussed in \cite{paper1,paper2}, we find
\beq
  \begin{aligned}
    \frac{f_{1}^\textrm{loc}}{\Gamma} &= \frac{2}{v_\infty^2} + \textcolor{1PN}{4}\label{floc}\,,\\
    \frac{f_{2}^\textrm{loc}}{\Gamma} &= \textcolor{1PN}{\frac{6}{v_\infty^2}} +\textcolor{2PN}{\frac{15}{2}}\,,\\
    \frac{f_{3}^\textrm{loc}}{\Gamma} &=
    \textcolor{2PN}{\left[-5\nu+\frac{17}{2}\right]\frac{1}{v_{\infty }^2}}
    +\textcolor{3PN}{\left[\frac{3 \nu ^2}{4}-\frac{37 \nu }{2}+9\right]}
    +\textcolor{4PN}{\left[-\frac{3 \nu ^3}{8}+\frac{33 \nu ^2}{16}-\frac{471 \nu }{80}\right] v_{\infty }^2}\\
    \frac{f_{4}^\textrm{loc}}{\Gamma} &=
    \textcolor{3PN}{\left[\frac{7 \nu ^2}{2}+\left(\frac{41 \pi ^2}{32}-\frac{160}{3}\right) \nu +8\right]\frac{1}{v_{\infty }^2}}
    +\textcolor{4PN}{\left[-\frac{5 \nu ^3}{4}+\frac{49 \nu ^2}{2}+\left(\frac{33601 \pi ^2}{6144}-\frac{3877}{45}\right) \nu+\frac{129}{16}\right]}\\
    \frac{f_5^\textrm{loc}}{\Gamma} &=
    \textcolor{4PN}{\left[-\frac{9 \nu ^3}{4}+\left(\frac{2579}{24}-\frac{205 \pi ^2}{64}\right) \nu ^2+\left(\frac{14173 \pi^2}{6144}-\frac{3707}{360}\right) \nu +6\right]\frac{1}{v_{\infty }^2}}
  \end{aligned}
\eeq
for the $f_i$'s in the expansion of the center-of-mass momentum in \eqref{firsov1}. These values capture all the needed information to \textcolor{4PN}{4PN} order. For instance, the application of the B2B map requires the even coefficients of the scattering angle, which can be obtained directly from the $f_i$ coefficients \cite{paper1,paper2}
\beq
 \chi_{j,\textrm{loc}}^{(6)} = \frac{ 5\pi v_{\infty }^6}{16 \Gamma^6}
\left(\frac{ f_2^3}{2}+3  f_1 f_3 f_2+3   f_4 f_2+\frac{3 }{2} f_3^2+\frac{3}{2}  f_1^2 f_4+3  f_1 f_5+\frac{3  f_6}{2}\right)\,.
 \eeq
Due to the scaling $f_i \simeq 1/v_{\infty}^2$ \cite{paper1,paper2}, the PN contribution from  $f_6$ is subleading, such that 
\begin{align}
 \chi_{j,\textrm{loc}}^{(6)} &= \textcolor{3PN}{\left[\frac{105 \pi  \nu ^2}{16} +\frac{5}{256} \pi  \left(123 \pi ^2-8000\right) \nu +\frac{1155 \pi }{8}\right]}\\   &\quad+\textcolor{4PN}{\left[-\frac{525 \pi  \nu ^3}{32}-\frac{5}{64} \pi  \left(123 \pi^2-7013\right) \nu ^2+\frac{\pi  \left(771585 \pi ^2-37556864\right) \nu }{24576}+\frac{45045 \pi }{64}\right] v_\infty^2} \textcolor{5PN}{+ \cdots }\nn \end{align}
which agrees with the known value to \textcolor{4PN}{4PN}. The same reasoning applies to higher order terms. It is now straightforward to construct the (reduced) local-in-time bound radial action following the B2B dictionary, which takes the form
\begin{equation}
  i_r^{\rm loc} \equiv \frac{\cS_{r, \rm ell}^{\rm loc}}{G M^2 \nu} = \textrm{sgn}(\hat{p}_\infty)\chi_j^{(1)}-j\left(1+\frac{2}{\pi}\sum_{n=1}^\infty\frac{\chi_{j, \rm loc}^{(2n)}}{(1-2n)j^{2n}}\right)\,,\label{ir}
\end{equation}
with the \textcolor{4PN}{4PN} contribution to the $\chi_{j, \rm loc}^{(2n)}$'s, up to $n=4$, obtained from the $f_i$'s. Using \cite{paper1,paper2}
\beq
v_\infty^2 \to \left(1+\cE + \frac{\nu\cE^2}{2}\right)^2-1\,,
\eeq
the bound radial action then becomes a function of the (negative) binding energy after analytic continuation. From the radial action we compute all the local-in-time observables for bound orbits through derivatives w.r.t. the angular momentum and binding energy. For example, we derived the local correction to the periastron advance to \textcolor{4PN}{4PN} via 
\beq 1+\frac{\Delta \Phi_{\rm loc}}{2\pi} = -{\partial \over \partial j} i_r^{\rm loc}\,,\eeq 
also directly from the value of the scattering angle using \eqref{Phij}, yielding perfect agreement with the result reported in \cite{Bernard:2016wrg}. The map can also be shown to hold for known values at higher PN orders. For example, the bound radial action constructed through \eqref{floc} neatly reproduces all the values presented in TABLE XIV of \cite{binidam2}. Furthermore, the B2B map between angle and periastron advanced was also used in the recent results at \textcolor{5PN}{5PN} in \cite{5pnfin2}.

\subsection{Large-$J$ expansion}
Here we explicitly check the validity of the B2B dictionary in the large-eccentricity limit for the paradigmatic example at \textcolor{4PN}{4PN} order. To evaluate the non-local dynamics will we use a (comprehensive) parameterization of the Newtonian orbit,
\begin{subequations}
\begin{align}
r&=\frac{j^2}{1+e\,\cos\alpha}\,,\\
\phi-\phi_0&=\alpha\,,\\
j^{-3}\,(t-t_0)&=\frac{2}{(1-e^2)^{3/2}}\arctan\left(\sqrt{\frac{1-e}{1+e}}\tan\frac{\alpha}{2}\right)-\frac{e}{1-e^2}\frac{\sin\alpha}{1+e\cos\alpha}\,,\\
\frac{dr}{dt}&=\frac{e\,\sin\alpha}{j}\,,\\
\frac{d\phi}{dt}&=\frac{(1+e\,\cos\alpha)^2}{j^3}\,.
\end{align}
\end{subequations}
After rescaling the time parameter $\tau \equiv \tfrac{t}{GM}$, the non-local Hamiltonian defined in \eqref{snloc} may be written as 
\begin{align}
H_\text{nloc}(\tau)=H_{\rm tail} + H_{\log r} \,,
\end{align}
with\footnote{The factor of $\log T$ is equivalent to the $\log (2e^{\gamma_E}\mu)$ in frequency space, see e.g. \cite{Damour:2014jta}.} 
\begin{align}
H_{\rm tail}&= -\frac{M \nu^2}{15j^{10}}{\rm Pf}_T\int\frac{d\tau'}{|\tau'-\tau|}F(\tau,\tau') \\
H_{\log r} &= -\frac{4M \nu^2}{15j^{10}} F(\tau,\tau)\log (r^2/T^2)\,,
\end{align}
where
$\text{Pf}_T$ stands for {\it partie finite}, and at leading PN order we have \cite{binidam1,binidam2}
\begin{align}
F(\tau,\tau')=(1+e\,\cos\alpha)^2\,(1+e\,\cos\alpha')^2\,(F_0+e\,F_1+e^2\,F_2)\,,
\end{align}
where
\begin{align}
F_0&=24\,\cos(2\alpha-2\alpha')\,,\\
F_1&=9\,\cos(2\alpha-3\alpha')+15\,\cos(\alpha-2\alpha')+9\,\cos(3\alpha-2\alpha')+15\,\cos(2\alpha-\alpha')\,,\notag\\
F_2&=-\frac{1}{4}\,\cos(\alpha+\alpha')+\frac{45}{8}\,\cos(3\alpha-\alpha')+\frac{45}{8}\,\cos(-3\alpha'+\alpha)\\ &+ \frac{27}{8}\,\cos(3\alpha-3\alpha')+\frac{77}{8}\cos(\alpha-\alpha')\,,\notag
\end{align}
with $\alpha\equiv \alpha(\tau)$ and $\alpha' \equiv \alpha(\tau')$. The non-local radial action is then obtained by integrating over the orbit,
\beq
\begin{aligned}
{\cal S}_r^{\rm nloc}=-\frac{GM}{2\pi}\int^{a}_{-a}d\alpha \frac{d\tau}{d\alpha} \big(H_\text{tail}+H_{\log r}\big) \equiv {\cal S}_r^{\rm tail} + {\cal S}_r^{\log r}\,,
\end{aligned}
\eeq
where the limits of integration given by hyperbolic-like ($a=\pi$) and elliptic-like ($a=\pi/2$) motion at leading order in $1/j$, respectively. Performing the (regularized) time and orbital integration we find
\beq
\begin{aligned}
\frac{{\cal S}_{r,\rm ell}^{\rm tail}}{GM^2\nu} &= 2\frac{{\cal S}_{r,\rm hyp}^{\rm tail}}{GM^2\nu}= \textcolor{4PN}{-\frac{1}{2\pi} \frac{16\pi \cE^2 \nu }{15 j^3} \left(37 \log \left(\frac{|\cE| T\,}{2j}\right)+100\right)}\,,\quad\quad\quad (\rm large\text{-}e\, approx.)\\
\frac{{\cal S}_{r,\rm ell}^{\log r}}{GM^2\nu} &= 2\frac{{\cal S}_{r,\rm hyp}^{\log r}}{GM^2\nu} = \textcolor{4PN}{-\frac{1}{2\pi}\frac{16\pi \,\nu\,\cE^2}{15j^{3}}\left(37\,\,\log \left(\frac{\sqrt{2}j}{T\,\sqrt{|\cE}|}\right)-\frac{85 }{4}\right)}\,,
\end{aligned}
\eeq
for each individual term, such that the sum becomes
\beq
\begin{aligned}
\frac{{\cal S}_{r,\rm ell}^{\rm nloc}}{GM^2\nu} = 2\frac{{\cal S}_{r,\rm hyp}^{\rm nloc}}{GM^2\nu}= \textcolor{4PN}{-\frac{1}{2\pi} \frac{16\pi \cE^2 \nu }{15 j^3} \left(37 \log \left(\frac{\sqrt{|\cE|}}{\sqrt{2}}\right)+\frac{315}{4}\right)}\,,\quad\quad\quad (\rm large\text{-}e\, approx.)
\end{aligned}
\eeq
which, at this order, formally obeys the B2B correspondence in \eqref{b2bnloc}. However, while the expansion of the radial actions in the limit of large angular momentum may be related through the B2B map, the non-local contributions obtained in this fashion do not fully describe generic bound orbits. Nevertheless, similarly to the local-in-time effects, we find the B2B dictionary correctly captures all the logarithms of the binding energy.

\subsection{Logarithms}

Let us consider now the logarithms (in binding energy) resulting from the non-local-in-time dynamics described by the universal contribution in \eqref{snloc}. As we mentioned earlier, the result takes the form,
\beq
\chi^{(n)}_{j, \rm nloc} = \chi^{(n)}_{j, \log} \log v_\infty + \cdots\,,
\eeq
which starts at \textcolor{4PN}{4PN} order, and has been computed to \textcolor{6PN}{6PN} in \cite{binidam2},\footnote{Except for $\chi_{j,\log}^{(8)}$, which we have obtained here through the Firsov representation.}
\beq
\begin{aligned}
   \label{chilog}    \frac{\chi_{j,\textrm{log}}^{(4)}}{\pi\nu} &=    -\textcolor{4PN}{\frac{37 v_{\infty }^4}{5}}   +\textcolor{5PN}{\left[\frac{111 \nu }{10}-\frac{1357}{280}\right] v_{\infty }^6}   +\textcolor{6PN}{\left[-\frac{111 \nu ^2}{8} +\frac{2517 \nu }{560}-\frac{27953}{3360}\right] v_{\infty }^8}+\cdots\,,\\ 
     \frac{\chi_{j,\log}^{(5)}}{\nu} &=
    -\textcolor{4PN}{\frac{6272 v_{\infty }^3}{45 }}
    +\textcolor{5PN}{\left[\frac{13952 \nu }{45}-\frac{74432}{525}\right] v_{\infty }^5}
    -\textcolor{6PN}{\left[\frac{21632 \nu ^2}{45 }-\frac{288224 \nu }{1575 }+\frac{881392}{11025 }\right] v_{\infty }^7}+\cdots\,,\\
\frac{\chi_{j,\log}^{(6)}}{\pi\nu} &=
   -\textcolor{4PN}{122  v_{\infty }^2}
   +\textcolor{5PN}{\left[\frac{811 \nu }{2}-\frac{13831 }{56}\right] v_{\infty }^4}
    +\textcolor{6PN}{\left[-785  \nu ^2+\frac{75595 \nu }{168}+\frac{64579 }{1008}\right] v_{\infty }^6}+\cdots\,,\\
           \frac{\chi_{j,\log}^{(7)}}{\nu} &=
    -\textcolor{4PN}{\frac{9344 v_{\infty }}{15}}
    +\textcolor{5PN}{\left[\frac{48256 \nu }{15}-\frac{284224}{105}\right] v_{\infty }^3}
   +\textcolor{6PN}{\left[-\frac{118912 \nu ^2}{15}+\frac{11456416 \nu }{1575}+\frac{587984}{567}\right] v_{\infty }^5}
    +\cdots\,,\\
 \frac{ \chi_{j,\log}^{(8)}}{\pi\nu} &=
  \textcolor{4PN}{
    -\frac{595}{3} }+\textcolor{5PN}{ \left[-\frac{15813 }{8}  + \frac{10535}{6} \nu \right]v_\infty^2}+\cdots
 \end{aligned}
 \eeq
 More recently, the logarithmic contribution to the scattering angle have been obtained at 4PM, to all orders in the PN expansion \cite{4pmeft,4pmtail}. The result reads 
\beq
\begin{aligned}
\frac{\chi^{(4)}_{j, \log}}{\nu\pi} &=  \frac{E}{ M^2\nu^2\pi}  {\partial \over \partial j} \Delta E_{\rm hyp}(j,\cE)  =   \frac{2 v_{\infty}^4}{\Gamma^3}   \chi_{2\epsilon}(\gamma)  \,,
 \end{aligned}
\eeq
with the explicit expression for $\chi_{2\epsilon}(\gamma)$ given in \cite{4pmeft,4pmtail}. Performing a PN expansion we find,
\begin{equation}
  \begin{aligned}
\frac{\chi^{(4)}_{j, \log}}{\nu\pi}    &= 
    \begin{multlined}[t]
      -\textcolor{4PN}{\frac{37 v_{\infty }^4}{5}}
      +\textcolor{5PN}{\left(\frac{111 \nu}{10}-\frac{1357}{280}\right) v_{\infty }^6}
      +\textcolor{6PN}{\left(-\frac{111 \nu ^2}{8}+\frac{2517 \nu }{560}-\frac{27953}{3360}\right) v_{\infty }^8}\\
    \textcolor{7PN}{+\left(\frac{259 \nu ^3}{16}-\frac{963 \nu ^2}{448}+\frac{2699 \nu }{224}+\frac{676273}{118272}\right) v_{\infty }^{10}}
      +\cdots\,,
    \end{multlined}
  \end{aligned}
\end{equation}
which neatly agrees with the PN derivations, and we added the first new result at \textcolor{7PN}{7PN} order. Due to the lack of information at higher PM orders, we will not consider this correction in what follows.\vskip 4pt

The Firsov representation can now be augmented to include the logarithmic corrections to describe generic orbits, re-written as
 \beq
 \bp^2 = p_\infty^2 \left[1 + \sum_i \left( f^{\rm loc}_i + f_i^{\log}\log |v_\infty|\right) \left(\frac{GM}{r}\right)^i\right]\,.\label{firlov}
 \eeq
 Notice that, as advertised, the factor of $\log r$ from potential-only contributions is traded by a logarithm of the binding energy in the full dynamics. Following the same steps as before we arrive at
 \beq
 \begin{aligned}
      \frac{f_4^{\log}}{\Gamma}\label{flog} &=
    -\textcolor{4PN}{\frac{296 \nu }{15}}
    -\textcolor{5PN}{\frac{1357 \nu  v_{\infty }^2}{105}}
    -\textcolor{6PN}{\frac{27953 \nu  v_{\infty }^4}{1260}}+\cdots\,,\\
   \frac{f_5^{\log}}{\Gamma} &=
    -\textcolor{4PN}{\frac{136 \nu }{3 v_{\infty }^2}}
    +\textcolor{5PN}{\left[\frac{796 \nu ^2}{15}+\frac{1271 \nu }{25}\right]}
    +\textcolor{6PN}{\left[-\frac{37 \nu ^3}{5}+\frac{50441 \nu ^2}{1050}+\frac{412281 \nu }{4900}\right] v_{\infty }^2}+\cdots\,,\\
     \frac{f_6^{\log}}{\Gamma} &=    \textcolor{5PN}{\left[\frac{2576 \nu ^2}{15}+\frac{5916 \nu }{25}\right]\frac{1}{v_{\infty }^2}}    
     -\textcolor{6PN}{\left[\frac{496 \nu ^3}{5}+\frac{453982 \nu ^2}{1575}-\frac{15073564 \nu }{33075}\right]}+\cdots\,,\\
     \frac{f_7^{\log}}{\Gamma} &= \textcolor{6PN}{\left[-348 \nu ^3-\frac{157946 \nu ^2}{105}+\frac{444883 \nu }{19845}\right]\frac{1}{v_{\infty }^2}}+\cdots\,,
\end{aligned}
\eeq
for the logarithmic contributions in the expansion of the center-of-mass momentum. Momentarily, we will show how the above coefficient reproduce the correct logarithmic contributions to the binding energy for quasi-circular orbits through the B2B dictionary.

\subsection{Large-eccentricity vs circular orbits}

In order to evaluate how well the bound radial action for generic orbits can be approximated by the B2B map from the scattering angle obtained in a large-eccentricity expansion, we consider next a paradigmatic example: the binding energy for circular orbits. Although, as we shall see, the B2B map does not reproduce to the exact value in the literature \cite{Damour:2014jta,Jaranowski:2015lha,Bernard:2017bvn,Marchand:2017pir,nrgr4pn1,nrgr4pn2}, we find that several contributions, notably local-in-time effects and leading tail logarithms, do transition smoothly between unbound and bound motion.\vskip 4pt

We start by constructing the B2B radial action from the knowledge of the total scattering angle, 
\begin{equation}
  i_r^{{\rm large\text{-}}e} = \textrm{sgn}(\hat{p}_\infty)\chi_j^{(1)}-j\left(1+\frac{2}{\pi}\sum_{n=1}^\infty\frac{\chi_{j, \rm loc}^{(2n)}+\chi_{j, \rm nloc}^{(2n)}}{(1-2n)j^{2n}}\right)\,,\label{irlarge}
\end{equation} 
including non-local-in-time effects,\footnote{In principle, to calculate the binding energy to \textcolor{4PN}{4PN} from the radial action we need the value of  $\chi_j^{(8)}$, which unfortunately is not presently known. Although we do not expect the Firsov representation to be valid for (non-logarithmic) non-local-in-time terms, we can still use it as a proxy for the full value. Hence, from
\begin{equation}
  \chi_j^{(8)} = \frac{35 \pi p_{\infty }^8}{256 \nu ^8 M^8}
  \begin{multlined}[t]
    \left(f_2^4+12 \left(f_1 f_3+f_4\right) f_2^2+12 \left(f_4 f_1^2+2 f_5 f_1+f_3^2+f_6\right) f_2+6 f_4^2\right.\\
    \left.+4 f_1^3 f_5+12 f_3 f_5+6 f_1^2 \left(f_3^2+2 f_6\right)+12 f_1 \left(2 f_3 f_4+f_7\right)+4 f_8\right)\,,
  \end{multlined}
\end{equation}
truncated to \textcolor{4PN}{4PN} order, we complete the expression in \eqref{irlarge}.} 
e.g. \cite{binidam2}
\beq
\begin{aligned}
\frac{\chi_{j,\textrm{nloc}}^{(4)}}{\pi\nu} &=
    \textcolor{4PN}{\left[-\frac{37}{5} \log  v_\infty + \frac{37}{5} \log 2 -\frac{63 }{4}\right] v_\infty^4} \textcolor{5PN}{+\cdots}\,,\\
  \frac{\chi_{j,\textrm{nloc}}^{(6)}}{\pi\nu} &=
    \textcolor{4PN}{\left[-122 \log v_\infty +122 \log 2 -\frac{99}{8} (21 \zeta_3+2)\right] v_\infty^2}\textcolor{5PN}{+\cdots}\,.
\end{aligned}
\eeq
The binding energy is then extracted by imposing the vanishing of the radial action to solve for the angular momentum as a function of binding energy, and subsequently using the first-law of binary dynamics \cite{letiec} to write $\cE$ as a function of the orbital frequency through the parameter $x \equiv (GM\Omega)^{2/3}$. To \textcolor{4PN}{4PN} order we~find~(recall $\epsilon=-2\cE$)
\begingroup
\allowdisplaybreaks
\begin{align}
  \epsilon(x) &=
  \left\{1
  +\left[-\frac{\nu }{12}-\frac{3}{4}\right] x
  +\left[-\frac{\nu ^2}{24}+\frac{19 \nu }{8}-\frac{27}{8}\right] x^2\right. \quad\quad \quad (\rm large\text{-}e\,\, approx.)\\
  &\quad+\left[-\frac{35 \nu ^3}{5184}-\frac{155 \nu ^2}{96}-\frac{5}{576} \left(246 \pi ^2-6889\right) \nu -\frac{675}{64}\right] x^3\nn\\
  &\quad+
  \begin{multlined}[t]
   \left[
      \frac{77 \nu ^4}{31104}+\frac{301 \nu ^3}{1728}+\frac{7 \left(2706 \pi ^2-71207\right) \nu ^2}{3456}+\frac{7 \left(19365 \pi ^2-98756\right)\nu }{23040}-\frac{3969}{128}\right.\nn\\  \left.
	  \left.+\textcolor{4PN}{\left(\frac{448 \log x}{15}-\frac{271768 \zeta_3}{45}+\frac{19576}{135}+\frac{463232 \log 2}{45}\right)\nu}
      \right]x^4\right\}+\textcolor{5PN}{\cdots}\,.
  \end{multlined}
\end{align}
For illustrative purposes we have colored here only the contribution from the non-local term, which enters at ${\cal O}(\nu)$.\vskip 4pt 

After direct comparison with the literature we find that all of the local-in-time contribution (and logarithms) perfectly matches the correct value for circular orbits. However, other terms, due to non-local effects, do not agree. At the end of the day there is a non-negligible mismatch, 
\beq
\delta \epsilon = \nu x^5 \frac{56}{135}   \Big(14559\, \zeta_3-329+144 \gamma_E - 24528 \log 2\Big) \simeq 10^2\, \nu x^5\,,
\eeq
between the correct result at this order and the one obtained from the B2B map using a large-eccentricity approximation. This feature (including the presence of different transcendental numbers!) remains essentially unaltered at higher PN orders: contribution from local-in-time effects transition smoothly to the correct result for generic orbits, whereas the numerical values for most of the non-local corrections translated from a large-eccentricity expansion through the B2B dictionary do not capture the circular case.\vskip 4pt  The situation, however, is remarkably different for the logarithmic terms. Using the expressions in \eqref{chilog}, or directly from the Firsov representation and the coefficients in \eqref{flog}, we obtain the following contribution to the bound radial action to \textcolor{6PN}{6PN} through the B2B map,
\begin{equation}
  i_r^{\log} = \frac{2}{\pi}\sum_{\textcolor{4PN}{n=2}}^{\textcolor{6PN}{n=4}}\frac{\chi_{j, \log}^{(2n)}}{(1-2n)j^{2n}}+\cdots\,.\label{irlog}
\end{equation} 
Performing the same manipulations as before, we find
\begin{equation}
  \begin{aligned}
    \epsilon_{\log x}&=\left\{ 
    \textcolor{4PN}{\frac{448}{15}} \nu  x^5
    +\left[\left(-\textcolor{4PN}{\frac{224}{5}}-\textcolor{5PN}{\frac{432}{5}}\right) \nu^2+\left(-\textcolor{4PN}{176}+\textcolor{5PN}{\frac{1172}{35}}\right) \nu \right]x^6 \quad\quad \quad (\rm large\text{-}e\,\, approx.)\right. \\
    &\quad +\left[\left(\textcolor{4PN}{\frac{616}{27}}+\textcolor{5PN}{\frac{792}{5}}+\textcolor{6PN}{\frac{176 }{3}}\right) \nu ^3+\left(\textcolor{4PN}{\frac{39776}{45}}+\textcolor{5PN}{\frac{491326}{315}}-\textcolor{6PN}{\frac{1394492}{945}}\right) \nu^2\right.\\
    &\quad\quad\quad \left.\left.+\left(-\textcolor{4PN}{\frac{2638064}{45}}+\textcolor{5PN}{\frac{6032774 }{105}}+\textcolor{6PN}{\frac{1138874}{1215}}\right) \nu \right]x^7\right\} \log x \,,
  \end{aligned}
\end{equation}
 for the logarithmic contribution to the binding energy obtained via scattering data to \textcolor{6PN}{6PN} order. (Notice this included different orders in the $1/j$ expansion of the scattering angle.)  After adding them all up, they beautifully reproduce the correct value for circular orbits. This demonstrates the validity of the B2B map and Firsov representation both for local-in-time as well as (leading) tail logarithms for generic orbits. 

\section{Discussion}\label{disc}

In this paper we have extended the B2B correspondence to include radiation effects, both in the dissipative and conservative sectors. We have found various types of analytic continuations between radiative observables for unbound and bound orbits, 
\beq
\Delta{\cal O}_{\rm ell} (J,\cE) = \Delta {\cal O}_{\rm hyp}(J,\cE) -\sigma_{\cal O} \Delta {\cal O}_{\rm hyp}(-J,\cE)\,,\nn
\eeq  
notably for the total (source) energy and angular momentum with $\sigma_{E/J} = +/-$, respectively; as well as conservative radiation-reaction contributions to the radial action,
\beq
\cS_{r, \rm ell}(J) = \cS_{r, \rm hyp}(J) - \cS_{r, \rm hyp}(-J)\nn\,,
\eeq
in the large-eccentricity limit. While the latter encapsulates all the local-in-time terms as well as the non-local-in-time logarithmic corrections, other non-local contributions in generic bound states are not captured by the large~$J$ expansion. This is easy to see already from the expression in \eqref{snloc}.
Assuming the validity of the adiabatic approximation, we can evaluate the integral over the conservative trajectory. Hence, performing a change of variables, e.g. $u \equiv  \omega \,e a ^{3/2}$, with $(e,a)$ the orbital elements  \cite{binidam1,binidam2}, as well as using the parameterized form for the trajectory in terms of $v$, the `eccentric anomaly,'
\beq
r_{\rm ell}(v) = a (1-e \cos v) \,, \quad\quad r_{\rm hyp}(v) = a (e \cosh v-1) \nn\,,
\eeq
 we can readily factor out the tail logarithms out of the integral, yielding a contribution to the radial action proportional to\footnote{More generally,  within the realm of the PM expansion, this is due to the fact that the pole from radiation modes in the tail effect is accompanied by a factor of $v_\infty^{-2(d-4)}$. The latter is ultimately responsible for the $\log v_\infty$ in the effective action~\cite{4pmeft,4pmtail}. Hence, the universal character of the divergent part of the tail implies that both appear multiplied by the total radiated energy at the end of the day, see \S\ref{seclog}.}
\beq
S^{\rm log}_{r, \rm ell/hyp} \propto \Delta E_{\rm ell/hyp} \log|\cE|\,.\nn
\eeq

As a result of the B2B map for the total radiated energy, this implies the B2B correspondence for the logarithmic terms in the conservative sector. However, the result also features left-over integrals of the sort (schematically)
\beq
\int_{\rm ell} F(u)\log u \, du \, \quad \quad {\rm vs} \quad\quad \int_{\rm hyp} F(u)\log u \, du\,,\nn \eeq
with $F(u)$ the (dimensionless) energy flux, see e.g. \cite{binidam1,binidam2} for examples in the PN context. 
As discussed in detail in \cite{binidam1,binidam2}, the presence of the $\log u$ produces wildly different values depending on the trajectory, involving even different types of transcendental numbers. For instance, while the quasi-circular binding energy features a factor of $\gamma_E$ at 4PN, the scattering angle does not. This precludes the existence of a straightforward map at the level of the final answer which would connect generic non-local-in-time contributions from the above integral evaluated on unbound and bound orbits.\vskip 4pt  It is possible, however, to establish a relationship at the level of the integrand. As it is well known, e.g. \cite{LucGer}, we can obtain the radial motion for elliptic-like orbits from hyperbolic-like motion directly via analytic continuation to negative binding energy and complex eccentric anomaly. Hence, we can search for the existence of an analytic continuation prior to performing the integration. For instance, on the one hand we find the hyperbolic result for the non-local-in time term at leading PN order leads to the integral 
\beq
\int_0^\infty dp\, p^6 | I_{\rm hyp}^{ij}(p)|^2 \log \left(p\,\Omega_{\rm hyp} \right)\,,\nn
\eeq 
where we made a change of variables $p \equiv   \,\omega/\Omega_{\rm hyp} $, and introduced a fictitious `orbital frequency' $\Omega_{\rm hyp} \equiv (2\cE)^{3/2}/GM$. Notice the parameter $p$ is a continuous variable in this case. The leading quadrupolar flux is given by $|I^{ij}_{\rm hyp}(p)|^2 = |\hat  I^{ij}_{\rm hyp}(i p)|^2 $,  as an analytic continuation of the following function, \begin{align}
&|\hat{I}^{ij}_{\rm hyp} (p)|^2=-\frac{8\, G^4 \,M^6\, \pi ^2 \,a^4 \,\nu ^2 }{3 \,e^4 \,p^4}\Bigg[3 e^2 \left(1-e^2\right) \Big(1+\left(1-e^2\right) p^2\Big) H_{p-1}^{(1)}(e p){}^2\notag\\
&-3 e \left(1-e^2\right) \Big(-e^2 p (2 p+3)+2 (p+1)^2\Big) H_p^{(1)}(e p) H_{p-1}^{(1)}(e p)\notag\\
&+\Big(-3 e^6 p^2+e^4 \left(12 p^2+9 p+1\right)-3 e^2 (p+1) (5 p+2)+6 (p+1)^2\Big) H_p^{(1)}(e p){}^2\Bigg]\nn\,,
\end{align}
with $(e,a)$ the orbital elements. The Hankel function is given by \beq
H_p^{(1)}(x) = \frac{1}{i\pi} \int_{-\infty}^{+\infty} dv \, e^{x \sinh v - p v} \,,\nn
\eeq 
which turns into the familiar $K_p(x)$ Bessel function after analytic continuation $x \to -i x$.\vskip 4pt On the other hand, the elliptic case involves a sum over harmonics of the type \cite{Damour:2015isa}
\beq
\sum_ {p=1}^{\infty} p^6 | I^{ij}_{\rm ell} (p)|^2 \log \left(p\,\Omega_{\rm ell} \right)\,,\nn
\eeq 
with $\Omega_{\rm ell} \equiv (-2\cE)^{3/2}/GM$ the true frequency of circular orbits (which may be obtained directly from $\Omega_{\rm hyp}$ via analytic continuation in $\cE$), and the quadrupole flux
\begin{align}
&|{I}^{ij}_{\rm ell} (p)|^2=\frac{8\, G^4 \,M^6\,a^4 \,\nu ^2 }{3 \,e^4 \,p^4}\Bigg[3 e^2 \left(1-e^2\right) \Big(1+\left(1-e^2\right) p^2\Big) J_{p-1}(e p){}^2\notag\\
&-3 e \left(1-e^2\right) \Big(-e^2 p (2 p+3)+2 (p+1)^2\Big) J_p(e p) J_{p-1}(e p)\notag\\
&+\Big(-3 e^6 p^2+e^4 \left(12 p^2+9 p+1\right)-3 e^2 (p+1) (5 p+2)+6 (p+1)^2\Big) J_p(e p){}^2\Bigg]\,,\nn
\end{align}
which is written in terms of the following Bessel function
\beq
J_p(x) = \frac{1}{2\pi} \int_{-\pi}^{+\pi} dv\, e^{i (x \sin v - p v)} \,. \nn
\eeq
The reader will immediately notice the resemblance between the two expressions. In particular, the analytic continuation in $p \to ip$ of the hyperbolic result plus the replacement $H^{(1)}_p \to J_p$ leads to the integrand, albeit with a discrete variable, for elliptic-like motion. Furthermore, the latter replacement between Bessel functions can be understood as the analytic continuation in $v \to iv$ associated with the link between unbound and bound (periodic) parameterizations of the orbit. From here we conclude that, indeed, we can find an analytic continuation at the level of the integrands. Yet, it is also clear the relationship ceases to apply once we expand in small/large eccentricity limits. As~a~consequence, other than the results uncovered here, it seems unlikely that a B2B-type map can connect non-local-in-time scattering data to generic bound orbits. It is still unclear, however, whether bound states with non-negligible eccentricities may be well approximated by the original B2B dictionary. We will explore this in more detail in future work.\vskip 4pt

We have concentrated here on local- and non-local-in-time tail effects. For the sake of completeness, let us conclude with a few comments regarding non-linear conservative memory terms, which are expected to be indistinguishable from other local-in-time effects and therefore readily incorporated in the B2B dictionary. The leading memory contribution may be computed through the same topology as in Fig.~\ref{tail2} (but including the quadrupole coupling and the full radiative field in the middle line instead of the monopole term and quasi-static potential). Similarly to tail terms, the separation between dissipative and conservative effects must be performed at the level of the full equations of motion, which may be derived using the Keldysh-Schwinger formalism~\cite{chadRR,chadprl,tail}. Yet, at  the \textcolor{5PN}{5PN} order at which radiation-reaction memory corrections start to contribute, we also encounter other non-linear dissipative effects correcting the leading (Burke-Thorne) back-reaction force. The additional terms may be obtained via the EFT approach by including the (seagull-type) non-linear worldline coupling between the binary's quadrupole and the curvature tensor, see e.g. \cite{andirad2,5pnfin2}.\footnote{These should not be confused with effects that enter the dynamics at quadratic order in the leading back-reaction force. In the PM EFT language of \cite{pmeft}, the latter arise through iterations involving the deflection due to the leading dissipative effects inserted into the tree-level radiation-reaction effective action.} As it turns out, the extra terms also entail three quadrupole moments and an even number of time derivatives, as in the memory contribution. Therefore, they can mimic the scaling of time-symmetric conservative effects.  Furthermore, total time derivatives (known as `Schott terms') may not only be present in the balance equations, they may also remain once averaged over the orbital motion. Hence, one must exercise special care when separating the various pieces entering the dynamics through the product of (more than two) multipole moments. Although yielding somewhat subtle effects, all the conservative memory terms are purely local in time, and therefore are automatically included in the original B2B dictionary. We~will discuss these contributions elsewhere.\vskip 1cm

\section*{Acknowledgments}
This work was supported by the ERC Consolidator Grant  {\it Precision Gravity: From the LHC to LISA}, provided by the European Research Council (ERC) under the European Union's H2020 research and innovation programme, grant agreement No.\,817791. R.A.P. would like to thank the organizers of the workshop ``Gravitational scattering, inspiral, and radiation," at the Galileo Galilei Institute for Theoretical Physics in Florence, where the extension of the B2B correspondence to the radiative sector was presented for the first time.\footnote{\url{https://www.youtube.com/watch?v=WZ7MEYjvT8s&t=40s} }\\

{\it Note added}: While the results of this project were prepared for submission, the work~of~\cite{justinb2b} appeared which has some overlap with the analysis in \S\ref{sec:rad1}-\S\ref{sec:rad3} restricted to the case of spin-independent effects. \\

\bibliographystyle{JHEP}
\bibliography{part4}
\end{document}